\documentclass[preprint,aps,superscriptaddress,floats,showpacs]{revtex4}
\usepackage{amsmath}
\usepackage{amssymb}
\usepackage{epsfig}

\begin{document}
\title{Linear theory of nonlocal transport in a magnetized plasma}
\author{A. V. Brantov}
\affiliation{Theoretical Physics Institute, Department of Physics, University of
Alberta, Edmonton T6G 2J1, Alberta,  Canada}
\author{V. Yu. Bychenkov}
\affiliation{P. N. Lebedev Physics Institute, Russian Academy of Science, Leninskii
Prospect 53, Moscow 119991, Russia}\affiliation{Theoretical Physics Institute,
Department of Physics, University of Alberta, Edmonton T6G 2J1, Alberta,  Canada}
\author{W. Rozmus}
\affiliation{Theoretical Physics Institute, Department of Physics, University of
Alberta, Edmonton T6G 2J1, Alberta, Canada}
\author{C. E. Capjack}
\affiliation{Department of Electrical Engineering, University of Alberta, Edmonton
T6G 2J1, Alberta, Canada}
\author{R. Sydora}
\affiliation{Theoretical Physics Institute, Department of Physics, University of
Alberta, Edmonton T6G 2J1, Alberta,  Canada}
\date{\today}
\begin{abstract}
A system of nonlocal electron-transport equations for small perturbations in a magnetized plasma is derived using the systematic closure procedure of V. Yu. Bychenkov {\it et al}., Phys. Rev. Lett. {\bf 75}, 4405 (1995). Solution to the linearized kinetic equation with a Landau collision operator is obtained in the diffusive approximation. The Fourier components of the longitudinal, oblique, and transversal electron fluxes are found in an explicit form for quasistatic conditions in terms of the generalized forces: the gradients of density and temperature, and the electric field. The full set of nonlocal transport coefficients is given and discussed. Nonlocality of transport enhances electron fluxes across magnetic field above the values given by strongly collisional local theory. Dispersion and damping of magnetohydrodynamic waves in weakly collisional plasmas is discussed. Nonlocal transport theory is applied to the problem of temperature relaxation across the magnetic field in a laser hot spot.
\end{abstract}
\maketitle

\section{Introduction}

Nonlocal transport theories remain the subject of active studies in plasma physics especially in laser produced plasmas. The strong dependence of the collisional mean-free-path on the particle kinetic energy limits the validity of standard local transport relations to very long inhomogeneity scale lengths $L$. The value of $L$ has to be at least a hundred times longer than the electron-ion mean-free-path, $\lambda_{ei}$, for the Chapman-Enskog perturbative approach \cite{braginskii} to be valid \cite{bychenkov0}. This condition can be easily violated in most of the hot plasmas produced by high energy lasers. Likewise, many magnetized plasmas in laboratory \cite{marchand} and astrophysics require kinetic modeling because of the large values of $\lambda_{ei}/L$. In the presence of a strong magnetic field electron transport can be localized in the transverse direction by the short gyroradius, $\rho$. Therefore, often only nonlocal transport in response to longitudinal gradients is considered (cf. e.g. \cite{chang}). We will show that for an arbitrary direction of the gradient, longitudinal transport coefficients are affected by the magnetic field in the nonlocal transport regime. 

Nonlocality of thermal transport has been well documented in experiments. For example, no agreement was found between theory and experiment for time-resolved two-dimensional images of x-ray emissions \cite{montgomery,nicolai} until nonlocal electron transport was included. It is also clear that self-generated magnetic fields play an important role in the evolution of laser fusion plasmas \cite{glenzer}. These observations have motivated our current study, which also complements previous works on magnetized plasmas \cite{chang,snyder,silin} in the regime of weak collisionality.  

We have applied the method of Ref. \cite{bychenkov} to derive nonlocal closure relations for a plasma with a magnetic field. The result is equivalent to the linearized kinetic theory of small amplitude perturbations. Our derivation involves a solution to the initial value problem for the electron kinetic equation with electron-electron (e-e) and electron-ion (e-i) Landau collision operators. The linearized kinetic equation is solved in the Fourier ${\bf k}$-vector space for slowly varying processes. We use approximations which make our results correct for high-$Z$ plasmas. We have neglected e-e collisions in the equation for the anisotropic part of the electron distribution function. Electron-electron collisions are only kept in the equation for the isotropic part of the distribution function which is unaffected by e-i collisions. In addition, the anisotropic part of the electron distribution function (EDF) is approximated by a single angular harmonic. This corresponds to the so called linearized diffusive Fokker-Planck model \cite{bernstein,albritton,brunner}. The original study \cite{bychenkov} of an unmagnetized plasma employed a summation technique \cite{luciani,epperlein}, which included an infinite set of angular harmonics and thus ensured validity of the transport theory over the entire range of particle collisionality including collisionless limit. Here, because of the geometrical complexity introduced by the magnetic field we are unable to carry the summation over the infinite set of harmonics. Therefore our results are applicable to an inhomogeneity scale length corresponding to ${\bf k}$-vectors such that $k_\perp \rho < 1$ and $k_\parallel \lambda_{ei} < 1$, where $k_\perp$ and $k_\parallel$ are wave vector components with respect to the direction of the magnetic field. However, in the case where the spatial gradients are almost orthogonal to the direction of magnetic field our theory is valid even for $k \lambda_{ei} >1$ provided $k \lambda_{ei} < \lambda_{ei}/\rho$. The validity condition $k_\parallel \lambda_{ei} < 1$ is more relevant to dense plasmas as for example in inertial confinement fusion targets. Finally the closure relations derived in this paper apply to small amplitude perturbations of temperature, density and electric field about the homogeneous equilibrium state. A rigorous generalization of these results to nonlocal and nonlinear transport theory is a difficult problem. One possible solution, \cite{batishchev} is based on linear theory results for varying in space background parameters. One can also gain a valuable insight into validity of different phenomenological models reproducing Fokker-Planck simulation results as in Ref. \cite{nicolai}.

An important difference between the nonlocal theory of unmagnetized plasmas \cite{bychenkov} and that for magnetized plasmas is that one now deals with three different components of the electron flux: longitudinal, transverse, and oblique. All of these are calculated here in the Fourier-space and are related by closure relations to perturbations in the temperature, density and electric field. Hydrodynamical variables have been defined in such a way that in strongly collisional limit we recover classical local results of the Chapman-Enskog method \cite{braginskii} .

The paper is organized as follows. Section \ref{sec2} presents a linearized solution to the kinetic FP equation for the EDF in a magnetized plasma. In Sec. \ref{sec3} we use this solution to derive the nonlocal electron flux. The strongly collisional limit is examined in Sec. \ref{sec4}. Section \ref{sec5} is devoted to study of nonlocal transport coefficients and their dependence on the collisionality parameter and the Hall parameter. Nonlocal hydrodynamic equations for small perturbations are formulated in Sec. \ref{sec6} and are used to derive dispersion relations and damping of MHD waves. In Sec. \ref{sec7} we apply our transport theory to study of the hot spot temperature relaxation in the context of laser produced plasma with a self-generated magnetic field. We conclude with a discussion and summary in Sec. \ref{sec8}.

\section{Electron distribution function in a magnetized plasma}\label{sec2}

We start from the standard electron kinetic equation for the EDF, $f_e$, in a high-Z plasma placed in a homogeneous magnetic field ${\bf B}$
\begin{equation}\label{eqfe}
\frac{\partial f_e}{\partial t}+ {\bf v} \cdot \frac{\partial f_e}{\partial {\bf r}} -
\frac{e}{m_e} \left( {\bf E}+\frac{1}{c} [{\bf v} \times {\bf B}] \right) \cdot \frac{\partial
f_e}{\partial {\bf v}} = C_{ei}[f_e]+C_{ee}[f_e,f_e]\,,
\end{equation}
where ${\bf E}$ is the electric field, $c$ is the speed of light, $e$ and $m_e$ are the electron charge and electron mass, $C_{ei}[f_e]$ and $C_{ee}[f_e,f_e]$ are electron–-ion and electron–-electron collision terms \cite{shkarofsky}.

Following the previous study \cite{bychenkov}, we develop a new nonlocal transport theory of a magnetized plasma which describes the plasma response to small--amplitude perturbations. The EDF, $f_e$, is written as the sum of Maxwellian function, $F_0$, (with homogeneous background electron density $n_e$ and temperature $T_e$) and an isotropic perturbation, $\delta f_0$, and an anisotropic perturbation, $\delta f_{an}$: $f_e = F_0 +\delta f_0(v) + \delta f_{an}({\bf v})$. Small amplitude linear perturbations are described in terms of Fourier components characterized by the wave vector ${\bf k}$. The linearized kinetic equation (\ref{eqfe}) gives two coupled equations for the perturbed EDFs, $\delta f_0$ and $\delta f_{an}$, in ${\bf k}$-space
\begin{eqnarray}
\label{system1} && \frac{\partial \delta f_0}{\partial t}+ i < {\bf k} \cdot {\bf v} f_{an} > -\frac{i}{3}({\bf k} \cdot {\bf u}) v \frac{\partial F_0}{\partial v}  = C_{ee}[\delta f_0, F_0]; \\ \label{system2} &&
\frac{\partial \delta f_{an}}{\partial t}+ [{\bf \Omega} \times {\bf v}] \frac{\partial \delta
f_{an}}{\partial {\bf v}}+ i {\bf k} \cdot {\bf v} f_{an} - i < {{\bf k} \cdot {\bf v}} f_{an} > - C_{ei}[f_{an}]
 = \\ &&- i {\bf k} \cdot {\bf v} f_0 + \frac {e \widetilde{\bf E}}{m_e} \frac{\partial F_0}{\partial {\bf v}} + i
\left[({\bf k} \cdot {\bf v })({\bf u} \cdot {\bf v})-\frac{1}{3}({\bf k} \cdot {\bf u }) v^2 \right] \frac{1}{v}
\frac{\partial F_0}{\partial v}\,, \nonumber
\end{eqnarray}
where ${\bf u}$ is the velocity of an ion flow, $\Omega = e B / m_e c$ is the electron gyrofrequency, brackets $<...>$ denote the averaging over angles defining the orientation of the velocity vector ${\bf v}$ and $\widetilde{ \bf E}={\bf E}+ {\bf u \times B}/c$. We have neglected electron-electron collisions in Eq. (\ref{system2}) because they play a smaller role compared to electron-ion collisions in the evolution of an anisotropic distribution function in a high-Z plasma.  

We introduce a unit vector ${\bf b} ={\bf B} /B$ in the direction of the magnetic field. Three components of a vector ${\bf A}$ can be defined with respect to ${\bf b}$ in the following manner: ${\bf A}_\parallel ={\bf b}({\bf A}\cdot{\bf b})$, ${\bf A}_\wedge={\bf b}\times{\bf A}$, and ${\bf A}_\perp={\bf b}\times[{\bf A}\times{\bf b}]$. Assuming long wavelength perturbations such that ${\bf k_{\perp} \cdot v} \ll \max({\Omega, \nu_{ei}})$, $ {\bf k}_{\parallel} \cdot {\bf v} \ll \nu_{ei}$ and slow time variations $\partial/\partial t \ll \Omega $ the solution to Eq. (\ref{system2}) for the anisotropic part of the EDF reads (cf. Ref. \cite{silin})
\begin{eqnarray}
&& \delta f_{an}  = -i \delta f_0 \left [ \frac {{k}_\parallel {v}_\parallel} {\nu_{ei}}
+\Omega \frac{{\bf k}_\wedge \cdot {\bf v}}{\nu_{ei}^2+\Omega^2}+\nu_{ei} \frac{{\bf k}_\perp \cdot {\bf v}}{\nu_{ei}^2+\Omega^2}\right ] -  \nonumber \\ && \frac{e F_0}{T_e} \left [ \frac {\widetilde{E}_\parallel {v}_\parallel} {\nu_{ei}} +\Omega \frac{{\widetilde{\bf E}}_\wedge \cdot {\bf v}}{\nu_{ei}^2+\Omega^2}+\nu_{ei} \frac{\widetilde{\bf E}_\perp \cdot {\bf v}} {\nu_{ei}^2+\Omega^2}\right ]-  \nonumber \\ && \frac{i F_0}{v_{Te}^2} \left[\frac{1}{3 \nu_{ei}} \left( ({\bf k}_\parallel \cdot {\bf v})({\bf u}_\parallel \cdot {\bf v}) + ({\bf k}_\perp \cdot {\bf v})({\bf u}_\perp \cdot {\bf v})- \frac{v^2}{3}{\bf k}\cdot{\bf u} \right )+ \right .
 \label{anis} \\
 && \left.  {v}_\parallel
 \frac{3 \nu_{ei} ({u}_\parallel ({\bf k}_\perp \cdot {\bf v}) + {k}_\parallel({\bf u}_\perp \cdot {\bf v}))-\Omega ({u}_\parallel({\bf k}_\wedge \cdot {\bf v})+ {k}_\parallel ({\bf u}_\wedge \cdot {\bf v}))} {9 \nu_{ei}^2+ \Omega^2}+ \right .
 \nonumber \\&& \left.
\frac{3 \nu_{ei} \Omega (({\bf k}_\perp \cdot {\bf v})({\bf u}_\perp \cdot {\bf v})-({\bf k}_\wedge \cdot
{\bf v})({\bf u}_\wedge \cdot {\bf v})) + 2 \Omega^2 (({\bf k}_\perp \cdot {\bf v})({\bf u}_\wedge \cdot
{\bf v})+ ({\bf k}_\wedge \cdot {\bf v})({\bf u}_\perp \cdot {\bf v}))}{3 \nu_{ei}(9 \nu_{ei}^2+4 \Omega^2)} \right ]\,, \nonumber
\end{eqnarray}
where $\nu_{ei}=4\pi Z n_e e^4\Lambda/m_e^2v^3$ is the velocity--dependent electron--ion collision frequency, and $\Lambda$ is the Coulomb logarithm. Expression (\ref{anis}) has been derived in the approximation, which corresponds to the so-called diffusive limit of the Fokker-Planck equation [cf. e.g. (\cite{brunner})]. This amounts to keeping only leading terms in the angular expansion of $\delta f_{an}$ as discussed in Ref. (\cite{silin}). Such approximation may always be justified by invoking a sufficiently large $Z$ and magnetic field values for the conditions $k_\perp Max\{\rho, \lambda_{ei}\}<1$ and $k_\parallel \lambda_{ei} < 1$ to be satisfied. By substituting $\delta f_{an}$ (\ref{anis}) into Eq. (\ref{system2}), we obtain an equation for the isotropic perturbation $\delta f_0$
\begin{eqnarray}\label{eqdeltaf_0}
&& \frac{\partial \delta f_0}{\partial t}+ \frac{v^2}{3} \left [ \frac{k_\parallel^2}{\nu_{ei}}+\frac{k_\perp^2 \nu_{ei}} {\nu_{ei}^2+\Omega^2} \right ] \left( \delta f_0 -\frac {i e ({\widetilde{\bf E}} \cdot {\bf k)} F_0}{k^2 T_e} \right )= \\ \nonumber && = C_{ee}[\delta f_0,F_0]- i {\bf k} \cdot {\bf u } \frac{m_e v^2}{3 T_e}F_0 +\frac{i e F_0 v^2\Omega}{3 T_e(\nu_{ei}^2+\Omega^2)}\left [ \frac{{\bf k}_\wedge \cdot [{\widetilde{\bf E}} \times {\bf k}]k_\parallel \Omega} {k^2 \nu_{ei}}+ {\bf k} \cdot {\widetilde{\bf E}}_\wedge  \right ]\,.
\end{eqnarray}
To derive nonlocal transport relations, we follow the method of Ref. \cite{bychenkov} and first solve an initial value problem for Eq. (\ref{eqdeltaf_0}). In a weakly collisional plasma the perturbation at $t=0$ provides a driver for transport processes. We postulate the initial $\delta f_0$ to be the linearized Maxwellian EDF, which is defined by density and temperature perturbations $\delta n_e\mid_{t=0}\equiv \delta n(0)$ and $\delta T_e\mid_{t=0}\equiv \delta T(0)$,
\begin{equation} \label{eqiM}
\delta f_0(v,0) = \left [ \frac{\delta n(0)}{n_e} + \frac{\delta T(0)}{T_e} \left(\frac{v^2}{2 v^2_{Te}} - \frac{3}{2} \right) \right] F_0(v)
\end{equation} 
With the initial condition (\ref{eqiM}), a solution to Eq. (\ref{eqdeltaf_0}) for $\delta f_0$ is expressed in terms of the basic functions $\Psi^A$:
\begin{eqnarray}\label{deltaf_0}
\frac {\delta f_0} {F_0}&=& \frac {i e ({\widetilde{\bf E}} \cdot {\bf k)}}{k^2
T_e}+\frac{\delta n(0)}{n_e} \Psi^N  +\frac{3}{2} \frac{\delta T(0)}{T_e} \Psi^T - i {\bf k} \cdot {\bf u} \Psi^u  +\\ \nonumber && +  \frac {i e {\bf k}_\wedge \cdot [{\widetilde{\bf E}} \times {\bf k}] k_\parallel} {k^4 T_e} \Psi^{E_1} + \frac {i e {\bf k} \cdot {\widetilde{\bf E}}_\wedge } {k^2 T_e} \Psi^{E_2}\, ,
\end{eqnarray}
where $\delta f_0$ (\ref{deltaf_0}) depends on four parameters: $\delta n(0)$, $\delta T(0)$, ${\bf u}$ and ${\widetilde{\bf E}}$. The ion flow velocity can be determined from a simple hydrodynamical model for cold ions and the field ${\widetilde{\bf E}}$ is usually evaluated from the quasineutrality condition. The solution given by Eq. (\ref{deltaf_0}) allows elimination of initial perturbations $\delta n(0)$, $\delta T(0)$ in terms of their current values by taking moments of $\delta f_0$. To do so and to proceed with closure relations for higher order moments we must find basic functions which satisfy the following equation with different source terms, $S_A$:
\begin{equation}\label{basic}
 \frac{v^2}{3}\left [\frac{k_\parallel^2}{\nu_{ei}}+\frac{k_\perp^2
 \nu_{ei}} {\nu_{ei}^2+\Omega^2} \right ] \Psi^A -
  \frac{1}{F_0} C_{ee}[\Psi^A,F_0]= S_A\, .
\end{equation}
Sources on the right hand side of Eq. (\ref{basic}) are defined as follows $S_N=1, S_T=v^2/3 v_{Te}^2-1$, $S_u= S_N + S_T$, $S_{E_1}= S_{\parallel}-S_{\perp}$, $S_{E_2}=S_{\wedge}$ and $S_a= k^2 v^2 / 3 \nu_a$, where the subscript $a$ assumes the values $\parallel$, $\wedge$, and $\perp$. The effective longitudinal and transverse collision frequencies read as follows
\begin{equation} \label{nua}
\nu_{\parallel} = \nu_{ei}\,, \quad
\nu_{\perp}= (\nu_{ei}^2+\Omega^2)/ \nu_{ei}\,, \quad \nu_{\wedge} =
(\nu_{ei}^2+\Omega^2)/ \Omega\,.
\end{equation}
Expressions (\ref{nua}) have been introduced following the standard definitions used in Ref. \cite{braginskii}.

As already stated, we can use Eq. (\ref{deltaf_0}) to calculate  hydrodynamic moments such as density and temperature perturbations 
\begin{eqnarray} &&\delta n_e=4\pi \int_0^\infty dv v^2 \delta f_0\,, \label{dndT} \\
&&\delta T_e=\frac{4\pi m_e}{3n_e} \int_0^\infty dv v^2 (v^2-3v_{Te}^2)\delta f_0\,.
\nonumber
\end{eqnarray}
and employ them to eliminate $\delta n(0)$ and $\delta T(0)$ from Eq. (\ref{deltaf_0}). As a result, the EDF is expressed in terms of its instantaneous hydrodynamic moments $\delta n_e$ and $\delta T_e$ and vectors ${\bf u}$ and ${\widetilde{\bf E}}$ as independent generalized thermodynamic forces. Moreover, because of the approximation $k_\parallel \lambda_{ei} \ll 1$ we have found that ion flow does not contribute to $\delta f_0$
explicitly. This has not been the case for short wavelength
perturbations, $k_\parallel \lambda_{ei} \gtrsim 1$ in studies of unmagnetized plasmas (cf. \cite{bychenkov}). The cancellation of terms containing ${\bf u}$ in the expression for $\delta f_0$ occurs because of the specific form of equation (\ref{basic}), which defines $\Psi^u$ as a linear combination of $\Psi^N$ and $\Psi^T$. Thus, the distribution function $\delta f_0$ can be expressed in terms of the velocity moments of the basic functions, $J_A^B = (4 \pi /n_e) \int_0^\infty d v v^2 \Psi^B S_A F_0$, as follows:
\begin{eqnarray}
&& \frac{\delta f_0 }{F_0} = \frac{i e ({\widetilde{\bf E}} \cdot {\bf k})} {k^2 T_e}
 + \frac{i e ({\widetilde{\bf E}^*} \cdot {\bf k})} {k^2 T_e} \frac{J_T^T
\Psi^N-J_T^N \Psi^T}{D_{NT}^{NT}} +\nonumber \\ && \frac {\delta T_e}{T_e}
\frac{(J_N^N+J_N^T) \Psi^T- (J_N^T+J_T^T) \Psi^N}{D_{NT}^{NT}} + \frac {i e {\bf k}_\wedge \cdot [{\widetilde{\bf E}} \times {\bf k}] k_\parallel}{k^4 T_e}\times \label{symmetr} \\ \nonumber && \left[ \Psi^{E_1} -
\frac{D_{NT}^{E_1T}}{D_{NT}^{NT}}\psi^N- \frac{D_{TN}^{E_1N}}{D_{NT}^{NT}}\psi^T
\right ] + \frac {i e {\bf k} \cdot {\widetilde{\bf E}}_\wedge }{k^2 T_e} \left[
\Psi^{E_2} - \frac{D_{NT}^{E_2T}}{D_{NT}^{NT}}\psi^N-
\frac{D_{TN}^{E_2N}}{D_{NT}^{NT}}\psi^T \right ]\,,\nonumber
\end{eqnarray}
where $D_{AB}^{CD}=J_A^C J_B^D-J_A^D J_B^C$ and ${\widetilde{\bf E}}^* =
{\widetilde{\bf E}} + i {\bf k} (T_e /e) ( \delta T_e/T_e  + \delta n_e/n_e) $ is the effective electric field which has been introduced in both, classical
\cite{braginskii} and nonlocal \cite{bychenkov} transport theories.

\section{Nonlocal electron fluxes}\label{sec3}

The expression for an anisotropic part of the EDF (\ref{anis}) is used to calculate electron fluxes: the electric current, ${\bf j}$, and the heat flux, ${\bf q}$,
\begin{eqnarray}\label{fluxes}
{\bf j} = - e \int d {\bf v} {\bf v} f_e\,,  \quad {\bf q} = - T_e \int d {\bf v} {\bf v}\left( \frac{5}{2}-\frac{v^2}{2 v_{Te}^2} \right) f_e\,.
\end{eqnarray}
Terms proportional to the ion velocity, ${\bf u}$, in Eq. (\ref{anis}) do not contribute to electron fluxes (\ref{fluxes}). This is due to the approximation $k_\parallel \lambda_{ei} \ll 1$. Equations (\ref{fluxes}) ignores in fact small contributions $\propto u \times O(k^2_\parallel \lambda^2_{ei})$.

Closure relations for the electric current density and the electron heat flux (\ref{fluxes}) have the same form as local transport relations \cite{braginskii} in strongly collisional plasmas. They involve linear combination of Fourier components of an effective electric field ${\bf E}^*$
and a temperature gradient $(\nabla T)_{\bf k}= i{\bf k}\delta T_e$,
\begin{eqnarray} \label{fluxes1}
{\bf j} &=& \mbox{\boldmath ${\sigma}$} \cdot \widetilde{\bf E}^*+ + i
\mbox{\boldmath $\alpha^j$}\cdot {\bf k} \delta T_e + \mbox{\boldmath
$\hat{\sigma}$} \cdot \widetilde{\bf E}  \,, \nonumber
\\{\bf q} &=& \mbox{\boldmath ${\alpha}^q$} \cdot \widetilde{\bf E}^*+ i {\bf k} \cdot \mbox{\boldmath $\chi$} \delta T_e + \mbox{\boldmath $\hat{\alpha}$} \cdot \widetilde{\bf E} \,.
\end{eqnarray}
However, the two transport matrices, $\mbox{\boldmath $\hat{\sigma}$}$ and $\mbox{\boldmath $\hat{\alpha}$}$ have no counterparts in the local relations \cite{braginskii}. Both coefficients vanish in the limit of $k \rightarrow 0$. The transport coefficients in (\ref{fluxes1}) are: electric conductivity tensors {\boldmath ${\bf \sigma}$}, thermoelectric tensors {\boldmath
$\alpha$} and temperature conductivity tensors {\boldmath $\chi$}. They are defined as follows ($c=j,q$)
\begin{eqnarray}
&&\!\!\!\!\!\! \mbox{\boldmath $\sigma$}= \!\left(\!
\begin{array}{ccc}\sigma_{\perp}& -\sigma_{\wedge}&0 \\
\sigma_{\wedge}&\sigma_{\perp}&0 \\ 0&0& \sigma_{\parallel}
\end{array}\!\right) \, \mbox{\boldmath $\alpha^c$}= \! \left (\!
\begin{array}{ccc}\alpha_{\perp}^c&-\alpha_{\wedge}^c&0 \\
\alpha_{\wedge}^c&\alpha_{\perp}^c&0 \\ 0&0& \alpha_{\parallel}^c \end{array}
\! \right )\, \, \mbox{\boldmath $\chi$}=\! \left(\!
\begin{array}{ccc}\chi_{\perp}&-\chi_{\wedge}&0 \\ \chi_{\wedge}&\chi_{\perp}&0 \\
0&0& \chi_{\parallel} \end{array} \! \right)
  \\ \nonumber
&&\!\!\!\!\!\!\!\mbox{\boldmath $\hat{\sigma}$}=\! \left(\!\!
\begin{array}{ccc}\!\!\sigma_{11} \displaystyle{\frac{k_{\parallel}^2}{k^2}}
&-\sigma_{12}&-\sigma_{11} \displaystyle{\frac{k_{\parallel}k_{\perp}}{k^2}} \\
\sigma_{21}
 \displaystyle{\frac{k_{\parallel}^2}{k^2}} & \sigma_{22}&-\sigma_{21}
 \displaystyle{\frac{k_{\parallel}k_{\perp}}{k^2}} \\ -\sigma_{11}
 \displaystyle{\frac{k_{\parallel}k_{\perp}}{k^2}}& \sigma_{12}
 \displaystyle{\frac{k_{\perp}}{k}}& \!\!\!\!\!\sigma_{11}
 \displaystyle{\frac{k_{\perp}^2}{k^2}} \end{array}\!\!\!\! \right )
\mbox{\boldmath $\hat{\alpha}$}=\!\! \left(\!\!\!\! \begin{array}{ccc}\,\, \alpha_{11}
\displaystyle{\frac{k_{\parallel}^2}{k^2}} &-\alpha_{12}&-\alpha_{11}
\displaystyle{\frac{k_{\parallel}k_{\perp}}{k^2}} \\
\!\!\!\!\!\!\! \!\! \alpha_{21}
 \displaystyle{\frac{k_{\parallel}^2}{k^2}} & \alpha_{22}&-\alpha_{21}
 \displaystyle{\frac{k_{\parallel}k_{\perp}}{k^2}} \\ -\alpha_{31}
 \displaystyle{\frac{k_{\parallel}
 k_{\perp}}{k^2}}& \alpha_{32} \displaystyle{\frac{k_{\perp}}{k}}& \!\!\!\!\!\!\alpha_{31}
 \displaystyle{\frac{k_{\perp}^2}{k^2}} \end{array}\!\!\!\! \right ) \end{eqnarray}
where ${\bf k} = (k_{\perp},0,k_{\parallel})$. Explicit expressions for these coefficients are given in Appendix A. All nonlocal transport coefficients
(\ref{tr_coef}) are functions of the wave vector and can be calculated by using solutions to the equations for the basic functions (\ref{basic}).

Equations (\ref{fluxes1}) represent the main results of our theory. They define constitutive relations for electron fluxes in terms of thermodynamic forces. Transport relations (\ref{fluxes1}) are linear in the temperature gradient and in a generalized field, similarly to local relations of the collision dominated hydrodynamics \cite{braginskii}. The nonlocality of expressions (\ref{fluxes1}) is displayed in terms of convolutions in the configurational space involving temperature gradient and generalized electric field, and kernels which are defined by the set of nonlocal transport coefficients. This new form of transport theory generalizes previous results \cite{bychenkov} by including magnetic field effects. The interplay between weakly collisional and magnetic field effects on transport theory is still poorly understood. However it can play an important role in the hydrodynamical evolution of laser produced plasmas as was demonstrated by the numerical simulations in Ref. \cite{nicolai}.

The transport coefficients (\ref{tr_coef}) depend on the following set of parameters: $k_\parallel \lambda_{ei}$, $k_{\perp} \rho$, $\Omega/\nu_{ei}^T$, and $Z$, where $ \rho=v_{Te}/\Omega$ is the electron Larmor radius and $\nu_{ei}^T=\sqrt{2/9\pi}\,\nu_{ei}(v_{Te})\equiv v_{Te}/\lambda_{ei}$ is
the electron--ion collision frequency for thermal electrons \cite{braginskii}. In Sec. \ref{sec5} the equivalent set of independent parameters: $k \lambda_{ei}$ (nonlocality parameter), $\Omega/\nu_{ei}^T$ (Hall parameter), $k_\perp/k_\parallel$, and $Z$ (ionic charge) will be used.

\section{Transport theory of a strongly collisional plasma}\label{sec4}

As a benchmark test for our transport theory we first consider the limit of strong collisions, i.e. $Z k^2 \lambda^2_{ei} \ll 1 $.  In this limit the basic functions $\Psi^A$ $(A=N,T,u,E_1,E_2)$ have the following form
\begin{equation}\label{psi}
\Psi^A = \frac{\lambda_{ei}}{v_{Te}} \left[c_0^A + c_1^A \left (\frac{v^2}{3
v_{Te}^2} - 1 \right) \right]\,.
\end{equation}
Expression (\ref{psi}) corresponds to a two polynomial approximation to the Laguerre polynomial series expansion of the symmetrical part of the EDF \cite{bychenkov0}. Coefficients $c_0^A$ and $c_1^A$ are found from the solution to Eq. (\ref{basic}) in the Lorentz plasma approximation.
From the moments of the EDF, $J_A^B$, we can evaluate transport coefficients (\ref{tr_coef}),
\begin{eqnarray}\label{eq}
&& \sigma_a =\sigma_a^0 =\frac{4 \pi e^2}{k^2 T_e}\int_0^\infty {dv v^2} S_a F_0 \,\,, \quad \alpha_a^j = \alpha_a^q = \alpha_a^0=\frac{4 \pi e}{k^2 T_e}\int_0^\infty {dv v^2} S_{q_a} F_0  \,, \nonumber \\ && \chi_a = \frac{4 \pi}{k^2}\int_0^\infty {dv v^2}S_{q_a}\left(\frac{v^2}{2 v_{Te}^2} -\frac{5}{2}\right) F_0\,,\quad \mbox{\boldmath $\hat{\sigma}$}=0\,, \quad \mbox{\boldmath $\hat{\alpha}$} = 0\,,
\label{Braginskii}
\end{eqnarray}
where $S_{q_a}=k^2 v^2 (v^2/v_{Te}^2-5)/6\nu_a$ ($a = \parallel$, $\wedge$, and $\perp$) and the effective collision frequencies, $\nu_a$, are given by Eq. (\ref{nua}). With expressions (\ref{Braginskii}), the electron fluxes (\ref{fluxes1}) assume the standard form:
\begin{eqnarray}
&&{\bf j} = \sum_a \sigma_a {\widetilde{\bf E}}_a + \sum_a {\alpha}_a i {\bf k}_a
\delta T_e  \,,\label{jflux}\\ && {\bf q} = - \sum_a {\alpha}_a T_e {\widetilde{\bf
E}}^*_a - \sum_a {\chi}_a i {\bf k}_a \delta T_e \,. \label{qflux}
\end{eqnarray}
As expected, longitudinal transport coefficients ($a=\parallel$) do not
depend on the magnetic field and correspond to the Spitzer--H$\ddot{a}$rm (SH)
result without magnetic field.

In the absence of a magnetic field ($\Omega \to 0$) all transverse and longitudinal transport coefficients are equal. The oblique coefficients are proportional to $\Omega$, i.e.
\begin{equation}
\sigma^0_{\wedge}= \frac{70 e^2 n_e \Omega}{\pi m_e {\nu_{ei}^{T}}^2}\,, \quad
\alpha^0_{\wedge}= - \frac{210 e n_e \Omega}{\pi m_e {\nu_{ei}^{T}}^2}\,, \quad
\chi_{\wedge}= \frac{1015 n_e T_e \Omega}{\pi m_e {\nu_{ei}^{T}}^2}\,.
\end{equation}
In the limit of a strong magnetic field $\Omega \to \infty$, transverse and oblique transport coefficients behave as $\propto 1/\Omega^2$ and $\propto 1/\Omega$, correspondingly,
\begin{eqnarray}&& \sigma^0_{\perp}= \frac{e^2 n_e \nu_{ei}^T}{m_e \Omega^2}\,\,, \quad
\alpha^0_{\perp}= \frac{3 e n_e \nu_{ei}^T}{2 m_e \Omega^2}\,\,, \quad \chi_{\perp}=
\frac{13 n_e T_e \nu_{ei}^T}{4 m_e \Omega^2}\,\,,\label{perp}\\ &&
\sigma^0_{\wedge}\!=\! \frac{e^2 n_e}{m_e \Omega}\left[1- \left( \frac{\pi^4
{\nu_{ei}^T}^5}{6 \Omega^5}\right)^{\!\!1/3}\!\right],\quad \alpha^0_{\wedge}\!=\! -
\frac{5 \pi e n_e}{2 m_e \Omega} \left( \frac{\pi {\nu_{ei}^T}^5}{6
\Omega^5}\right)^{1/3}\!\!\!\!\!\!, \quad \chi_{\wedge}= \frac{5 n_e T_e}{2 m_e
\Omega}\,\,, \label{wedge}
\end{eqnarray} 
with exception of the coefficient $\alpha^0_{\wedge}$, which varies as $\Omega^{-8/3}$. Note, that this fractional dependence \cite{epperlein1} on the electron gyrofrequency follows directly from the integral definition of the oblique thermocurrent coefficient in Eq. (\ref{eq}).

To simplify a comparison with results of classical transport theory \cite{braginskii,epperlein1} we rewrite the expression for an electron heat flux (\ref{qflux}) in terms of the electric current density instead of the effective electric field by using the following relation
\begin{eqnarray}\label{e_j}
&& {\bf E}^* = - \Omega m_e/(e^2 n_e) {\bf j}_{\wedge} + \sum_a \mu_a {\bf j}_a - \sum_a \beta^{uT}_a i {\bf k}_a \delta T_e \,.
\end{eqnarray}
Similarly, one can represent ${\bf q}$ in the following form
\begin{equation}\label{q_j}
{\bf q}= - \sum_a T_e \beta^{uT}_a {\bf j}_a - \sum_a \kappa_a i {\bf k}_a \delta T_e\,.
\end{equation}
The above expressions are now indentical to Ohm's (\ref{e_j}) and Fourier's (\ref{q_j}) laws as derived in Ref. \cite{braginskii} (the Braginskii's coefficient $\alpha_a$ is replaced here by $\mu_a$). Coefficients $\mu_a$, $\kappa_a$, and $\beta^{uT}_a$ in Eqs. (\ref{e_j}) and (\ref{q_j}) are defined as follows
\begin{eqnarray}\label{br}
&& \mu_{\parallel} = \frac{1}{\sigma^0_{\parallel}}\,, \quad \beta^{uT}_{\parallel} = \frac{\alpha^0_{\parallel}}{\sigma^0_{\parallel}}\,,
\quad \kappa_{\parallel}= \chi_{\parallel}-\frac{T_e
{\alpha^0_{\parallel}}^2}{\sigma^0_{\parallel}}\,, \\ \nonumber
&&\mu_{\perp} = \frac{\sigma^0_{\perp}} {{\sigma^0_{\perp}}^2+{\sigma^0_{\wedge}}^2}\,, \quad \mu_{\wedge} =\frac{\Omega m_e}{e^2 n_e}-\frac{\sigma^0_{\wedge}} {{\sigma^0_{\perp}}^2+{\sigma^0_{\wedge}}^2} \,,  \\
\nonumber &&  \beta^{uT}_{\perp} =
\frac{(\alpha^0_{\perp} \sigma^0_{\perp}+ \alpha^0_{\wedge} \sigma^0_{\wedge})}
{{\sigma^0_{\perp}}^2+{\sigma^0_{\wedge}}^2}\,, \quad \kappa_{\perp}=
\chi_{\perp}-\frac{T_e( {\alpha^0_{\perp}}^2 \sigma^0_{\perp}+2 \alpha^0_{\perp}
\alpha^0_{\wedge} \sigma^0_{\wedge}-{\alpha^0_{\wedge}}^2
\sigma^0_{\perp})}{{\sigma^0_{\perp}}^2+{\sigma^0_{\wedge}}^2}\,, \\ \nonumber &&
\beta^{uT}_{\wedge} = \frac{(\alpha^0_{\wedge} \sigma^0_{\perp}-\alpha^0_{\perp}
\sigma^0_{\wedge})} {{\sigma^0_{\perp}}^2+{\sigma^0_{\wedge}}^2}\,, \quad \kappa_{\wedge}=
\chi_{\wedge}-\frac{T_e( {\alpha^0_{\wedge}}^2 \sigma^0_{\wedge}+2 \alpha^0_{\perp}
\alpha^0_{\wedge} \sigma^0_{\perp}-{\alpha^0_{\perp}}^2
\sigma^0_{\wedge})}{{\sigma^0_{\perp}}^2+{\sigma^0_{\wedge}}^2}\,.
\end{eqnarray}
Transport coefficients given by (\ref{br}) are the same as the results of Ref. \cite{epperlein1}, where a detailed comparison with the strongly collisional results of Braginskii \cite{braginskii} was made. Reference \cite{epperlein1} also gives the functional dependence of $\beta^{Tu}_{\perp}$ and $\mu_{\wedge}$ on electron gyrofrequency $\Omega$.

\section{Nonlocal transport coefficients}\label{sec5}

In this section we will describe the dependence of the transport coefficients on the collisionality parameter, $k \lambda_{ei}$, for different values of the Hall parameter: $\Omega/\nu_{ei}^T =0.1 - 100$. We start with the longitudinal transport coefficients. In the local transport theory the longitudinal coefficients do not depend on magnetic field. This is not true in general for nonlocal transport, except for the special case when the direction of the plasma inhomogeneity is along the direction of the magnetic field ($k \equiv k_{\parallel}$). In the general case ($k \neq k_{\parallel}$) the dependence of transport coefficients, including longitudinal coefficients, on the magnetic field is defined by the symmetric part of the EDF which is a function of $k_{\perp} \rho$.

In the case of weak plasma inhomogeneity along the magnetic field, $k_{\perp}\gg k_{\parallel}$, transport coefficients $\sigma_\parallel$ and $\chi_\parallel$ normalized to their classical strongly collisional values are shown in Figs. \ref{fig1} and \ref{fig2} as functions of the collisionality parameter $k\lambda_{ei}$. The nonlocality of transport coefficients is characterized by very different functional dependence on $k\lambda_{ei}$ depending on the magnitude of the Hall parameter, $\Omega /\nu_{ei}^T$. For small values of the Hall parameter, $\Omega /\nu_{ei}^T < 1$, the transport coefficients steadily decrease with increasing $k\lambda_{ei}$. However, as shown in Figs. \ref{fig1} and \ref{fig2}, they can exceed the standard classical values at $k\lambda_{ei}\gtrsim 1$ for $\Omega /\nu_{ei}^T > 1$. Note, that the regime of $k\lambda_{ei}\gtrsim 1$ can still be within the validity condition of our theory, $k_\parallel\lambda_{ei}<1$, when the plasma gradients are almost transverse to the magnetic field, i.e. $k_\perp \gg k_\parallel$.

Transport in the case of a strong plasma inhomogeneity along magnetic field $k_{\perp}\ll k_{\parallel}$ is similar to the well known zero magnetic field case \cite{bychenkov}, therefore the next example considered is for $k_{\perp} \sim k_{\parallel}$. Figures \ref{fig3} and \ref{fig4} show transport coefficients $\sigma_\parallel$ and $\chi_\parallel$ as functions of $k\lambda_{ei}$ for $k_{\perp} = k_{\parallel}$. This corresponds to an angle of $45^\circ$ between the magnetic field vector and the direction of a linear gradient. As it is shown in Figs. \ref{fig3} and \ref{fig4} electrical and temperature conductivities increase with magnetic field reaching maximum values at $\Omega /\nu_{ei}^T \approx 0.3$ (cf. curves 3 in both Figs. \ref{fig3} and \ref{fig4}). Further increases in the parameter $\Omega /\nu_{ei}^T$ results in a small decrease in the coefficients $\sigma_\parallel$ and $\chi_\parallel$. At large values of $\Omega /\nu_{ei}^T$, they asymptotically approach curves 4 in Figs. \ref{fig3} and \ref{fig4}. In general, the longitudinal transport coefficients display smaller effects of nonlocality for larger magnetic field values.

Unlike the longitudinal coefficients, the perpendicular and oblique transport coefficients demonstrate strong dependence on magnetic field. To illustrate this, coefficients $\sigma_{\wedge,\perp}$ and $\chi_{\wedge,\perp}$ are shown in Figs. \ref{fig5} - \ref{fig8} as functions of collisionality parameter for different magnetic field values. In general, a strong magnetic field suppresses the effects of nonlocality. For instance, the transverse temperature conductivity, $\chi_{\perp}$, is practically independent of the collisionality parameter over the entire range of variations, $k\lambda_{ei}<1$, at $\Omega /\nu_{ei}^T=3$ (curve 5 in Fig. \ref{fig6}). As in the longitudinal transport case, the coefficients $\sigma_{\wedge,\perp}$ (Fig. \ref{fig5}) and $\alpha^q_{\wedge,\perp}$ increase slightly with the collisionality parameter at $\Omega /\nu_{ei}^T\gg 1$. One can also see that nonlocal effects are stronger for oblique transport coefficients as compared to transversal coefficients (c.f. Fig. \ref{fig5} and Fig. \ref{fig7}, and Fig. \ref{fig6} and Fig. \ref{fig8}).

Figure \ref{fig9} shows the ratio of the perpendicular to the longitudinal temperature conductivities as a function of Hall parameter. This ratio increases with the value of the nonlocality parameter, $k \lambda_{ei}$. Hence, in a plasma with steep gradients, the inhibition of electron fluxes due to a magnetic field is not as pronounced as in the strongly collisional case. This corresponds to an increase in the effective collision frequency (decrease of a mean free path) in the nonlocal limit \cite{bychenkov}. The kinetic effect of reducing the electron mean free path due to transport nonlocality was included in recent 2D hydrodynamical simulations \cite{nicolai} by replacing parameter $(\Omega /\nu_{ei}^T)^2$ with $d\,(\Omega /\nu_{ei}^T)^2$ where $d$ is a constant that should be defined from the solution to the FP equation. However, our results demonstrate that the phenomenological constant $d$ should be more correctly replaced by a function of the gradient scale length, $k \lambda_{ei}$.

To complete our discussion of nonlocal, magnetic field dependent transport coefficients, note that the analysis of $k \lambda_{ei}$ and $\Omega /\nu_{ei}^T$ dependencies can be performed for all other coefficients, $\alpha^c$, ${\bf \hat{\sigma}}$, and ${\bf \hat{\alpha}}$. These are explicitly calculated in Appendix A. As in the nonlocal transport case without a magnetic field \cite{bychenkov}, the coefficients $\alpha^c$ in the present theory may change signs at some wave numbers. The tensor coefficients, ${\bf \hat{\alpha}}$ and ${\bf \hat{\alpha}}$, are negligible as $\Omega /\nu_{ei}^T \to 0$, in agreement with the local transport theory \cite{braginskii}. They are also very small for the case of a strong magnetic field $\Omega /\nu_{ei}^T \to \infty$. This is the reason why these coefficients contribute to the electron fluxes only at $\Omega /\nu_{ei}^T\lesssim 1$. Finally, all conclusions reached in the above discussion are valid only in the limit $k_\parallel \lambda_{ei}< 1$ and $k_\perp < Max\{1/\rho, 1/\lambda_{ei}\}$. These are the required conditions for our theory to be valid.

\section{Magnetohydrodynamic waves}\label{sec6}

The first two velocity moments of the Eq. (\ref{system1}) yield continuity and energy balance equations for the electrons
\begin{eqnarray}\label{hydroe}
&& \frac{\partial n_e}{\partial t} + i n_e { \bf k} \cdot {\bf u} - \frac{1}{e} i
{\bf k} \cdot {\bf j} = 0\,,
\\ \nonumber
&& \frac {\partial T_e}{\partial t} +  \frac{2}{3 n_e} i {\bf k} \cdot {\bf q} -
\frac{2 T_e}{3 e n_e} i {\bf k} \cdot {\bf j}+ \frac{2 T_e}{3} i {\bf k} \cdot {\bf
u}=0\,.
\end{eqnarray}
Equations (\ref{hydroe}) and the hydrodynamic equation of motion for ions,
\begin{eqnarray}\label{hydroi}
\frac {\partial {\bf u}}{\partial t} = \frac{Z e}{m_i} {\widetilde{\bf
E}}+\frac{1}{n_i m_i} {\bf R}_{ie}\,,
\end{eqnarray}
together with nonlocal transport relations (\ref{fluxes1}) can be used in description of linear perturbations in a magnetized plasma. An ion-electron friction force, ${\bf R}_{ie}$, is defined by the anisotropic part of the EDF and can be expressed in terms of the electron current and the effective electric field,
\begin{equation}\label{Rei}
{\bf R}_{ie} = m_e \int d {\bf v v}\nu_{ei} f_e = -e n_e {\widetilde{\bf E}}^* - \frac{\Omega_e m_e}{e} {\bf j}_\wedge\,,
\end{equation}
and eventually in terms of the transport coefficients $\alpha^j$ and $\sigma$.
No other transport coefficients (c.f. Ref. \cite{bychenkov}) contribute to ${\bf R}_{ie}$ in the limit of $k\lambda_{ei}<1$.

By introducing perturbations of the form $\propto \exp(-i\omega t)$ into electron (\ref{hydroe}) and ion (\ref{hydroi}) fluid equations and by using Eq. (\ref{Rei}) we obtain
\begin{eqnarray}\label{hydro1}
&& e \omega \delta n_e + {\bf k} \cdot {\bf j} - e n_e {\bf k} \cdot {\bf u}=0
\nonumber \\  && \frac {3e}{2}n_e \omega \delta T_e - e {\bf k} \cdot {\bf q} + T_e
{\bf k} \cdot {\bf j} - e T_e n_e {\bf k} \cdot {\bf u} =0 \\ \nonumber && \omega
{\bf u}=   {\bf k} c_s^2 \left( \frac{\delta n_e}{n_e}+ \frac{\delta T_e}{T_e}
\right)- i \frac {\Omega_i}{e n_e} {\bf j}_{\wedge}\,,
\end{eqnarray}
where $\Omega_i= Z e B/m_i c$  is the ion gyrofrequency. These equations together with the closure relations (\ref{fluxes1}) and the standard wave equation, 
\begin{equation}\label{wave}{\bf k}^2 {\bf E} - ({\bf E} \cdot  {\bf k}){\bf k} - 4 \pi i \omega /c^2 {\bf j} =0\,, \end{equation} 
allow a study of linear modes of the magnetized plasma in the low frequency limit, $\omega \ll k c, \omega_{pe}$, where $\omega_{pe}$ is the electron plasma frequency.

The determinant of the system of equations (\ref{hydro1}) and (\ref{wave}) gives a dispersion relation for quasistatic magnetohydrodynamic (MHD) waves. It has been derived in the Appendix B in the following form
\begin{eqnarray}\label{deq}
&&A (3 \omega^2 \xi_A - 5 c_s^2 k_A^2) + B (3 \omega^2 - 5 k^2 c_s^2) +\nonumber \\&&C (\omega^2 - k c_s^2 ) + D  (\omega^2 \xi_A - c_s^2 k_A^2) + E \Omega_i \omega = 0\,,\\&& \xi_A = \frac{4 \pi}{c^2 k^2}(\omega^2 - k^2 v_A^2)\,, \qquad k_{A}^2 = \frac{4 \pi}{c^2}(\omega^2 - k_{\parallel}^2 v_A^2)\,,\nonumber
\end{eqnarray}
Symbols used in Eq. (\ref{deq}) are defined in Appendix B. In the strongly collisional limit, $k\lambda_{ei}\to 0$, the dispersion equation (\ref{deq}) has been studied in Ref. \cite{woods} in the isothermal approximation. In this approximation, long wave contributions given by coefficients $A$ and $B$ are neglected because they account for the adiabatic plasma response. Without a magnetic field, Eq. (\ref{deq}) corresponds to the ion acoustic dispersion relation (33) of Ref. \cite{myatt}.

Three MHD modes are defined by different branches of the solution to Eq. (\ref{deq}) for the real part of $\omega$. They are shown in Figs. \ref{fig10}a and \ref{fig11}a and correspond to the well known low frequency waves. The highest frequency branch (curves 1 in Figs. \ref{fig10}a and \ref{fig11}a) represents the fast MHD wave, $\omega=\omega_+$ where $\omega_\pm^2=(k^2(v_A^2+c_s^2)\pm \sqrt{k^4(v_A^2+c_s^2)^2-4k^2 k_\parallel^2v_A^2 c_s^2})/2$. The second branch (curve 2) includes, from small to large values of $k \lambda_{ei}$: the Alfven wave ($\omega=k_\parallel v_A\ll \Omega_i$), first ion cyclotron wave ($\omega=\Omega_i$) and ion acoustic wave ($\omega=kc_s$), respectively. Finally, the lowest frequency branch (curve 3) describes the slow MHD wave, $\omega=\omega_- \ll \Omega_i$ at small $k \lambda_{ei}$ and a second ion cyclotron wave ($\omega=\Omega_i k_\parallel/k$) for large $k \lambda_{ei}$.

Depending on plasma conditions, nonlocal effects may change the damping of MHD modes. This is shown in Figs. \ref{fig10}b and \ref{fig11}b for a propagation angle of 80$^\circ$ with respect to the direction of a magnetic field. The almost transverse direction of the vector ${\bf k}$ allows examination of the high $k \lambda_{ei}$ values in Figs. \ref{fig10} and \ref{fig11} within the applicability limits of our theory, $k_\parallel \lambda_{ei} \ll 1$. Two parameters, $p=(\omega_{pe} v_{Te}/ \nu^T_{ei}c)^2$ and $h^2=(\Omega /\nu^T_{ei})^2$, define different regimes of MHD wave damping. The $p/h^2$ is a well-known $\beta$-parameter, $\beta=c_s^2/v_A^2$. Figure \ref{fig10} corresponds to a high-$\beta$ (high pressure) plasma and Fig. \ref{fig11} describes a low-$\beta$ (low pressure) case. Note, that we do not discuss effects of ion viscosity. This can be included independently by using the standard approach of Ref. \cite{woods}.

In a high-$\beta$ plasma all modes except the fast MHD wave experience the usual collisional damping. The damping of the fast MHD wave in this case is unchanged up to the wavenumber $k\simeq 0.3/\lambda_{ei}$ and is enhanced at higher wavenumbers (Fig. \ref{fig10}b). It can reach values up to an order of magnitude higher than the damping rates given by the usual collisional approach \cite{woods} close to the short wave length applicability limit of our model. In a low-$\beta$ plasma, the slow MHD wave and the ion-acoustic wave can experience enhanced damping due to nonlocal effects (Fig. \ref{fig11}b) while the fast MHD mode has the usual damping. The enhanced damping of the slow MHD wave appears at $k> 0.1/\lambda_{ei}$ and remains up to $k\simeq 3/\lambda_{ei}$ until the slow MHD wave becomes a second ion cyclotron wave. At the same wavenumber, $k\simeq 3/\lambda_{ei}$, the ion-acoustic wave appears (the curve 2 in Fig. \ref{fig11}). The damping of ion-acoustic wave can be much higher than the standard collisional damping. This agrees with the result of Ref. \cite{bychenkov0} because of the negligible role that is played by the magnetic field in this case.

\section{Relaxation of hot spot temperature in a magnetic field.}\label{sec7}

Current laser plasma interaction experiments involve random phase plate laser beams which control the intensity distribution in the focal plane. This makes a hot spot geometry for the temperature inhomogeneity a basic element of transport theory in such plasmas. Typically the characteristic scales of the temperature gradient are in the regime of nonlocal transport. A self-generated magnetic field has also been well documented in such plasmas. Its existence motivates the discussion below, where we apply our transport theory to a single hot spot transport problem with magnetic field effects included.

We consider the relaxation of a single hot spot in a plasma with a given magnetic field. The geometry of our model involves an initial temperature perturbation with cylindrical symmetry which evolves due to thermal transport. The magnetic field is assumed orthogonal to the radial temperature gradient, i.e. it can have only $B_z$ or $B_\varphi$ components. We will examine conditions in laser produced plasmas where magnetic field effects influance transport in the nonlocal regime.  The solution of the initial value problem follows Ref. \cite{senecha}, where the relaxation of a hot spot was studied in an unmagnetized plasma. A temperature perturbation in a cylindrical hot spot is assumed to have a Gaussian profile
\begin{equation}
\delta T_0 (r) = \delta T_0 \exp \left( - \frac{r^2}{R^2} \right )
\end{equation}
at time, $t=0$. We also assume that the plasma flow has a negligible effect on the temperature relaxation, ${\bf u} = 0$.

Our model includes the energy balance equation (\ref{hydroe}) for electrons \begin{equation}\label{relax1}
\frac{\partial \delta T_e}{\partial t} = - i \frac{2}{3 n_e} {\bf k}_\perp \cdot {\bf q}_\perp\,,
\end{equation}
where the heat flux across magnetic field, ${\bf q}_\perp$,  (\ref{fluxes1}) has the following form ${\bf k}_\perp \cdot {\bf q}_\perp = - i {\bf k}_{\perp}^2
\kappa_{\perp} \delta T_e$. The value $\kappa_{\perp} = \chi_{\perp} - T_e
({\alpha^j_{\perp}})^2/\sigma_{\perp}$ is the transverse nonlocal heat
conductivity.

The solution to the equation (\ref{relax1}) is given by Eq. (5) in Ref. \cite{senecha}. The radial temperature and electron heat flux profiles are presented in Fig. \ref{fig12} for the magnetic field corresponding to Hall parameter $\Omega /\nu^T_{ei} = 0.3$. The result of the classical transport approach \cite{epperlein} (dashed line) overestimates the electron heat flux (Fig. \ref{fig12}b). Because of this reason, the hot spot temperature obtained from local theory \cite{epperlein} decreases faster (Fig. \ref{fig12}a) than predictions from Eq. (\ref{relax1}). As discussed in the previous section, a combination of the magnetic field and nonlocal effects leads to a slightly larger heat flux and smaller temperature as compared to the nonlocal and unmagnetized case. This is illustrated by dotted (present theory) and continuous lines (nonlocal and unmagnetized theory from Ref. \cite{senecha}) in Fig. \ref{fig12}. 

As expected, effects of nonlocal thermal transport become negligible with increasing values of the magnetic field. This is illustrated by Fig. \ref{fig13}, where the relaxation time (the time necessary for the temperature in the center of the hot spot to drop to half of its initial value) is shown as a function of the hot spot size, $R/\lambda_{ei}$ for different magnetic fields. For $R/\lambda_{ei} \gtrsim 3$ and $\Omega /\nu^T_{ei}= 1$ the effect of nonlocality almost vanishes. For small Hall parameters, $\Omega /\nu^T_{ei} \lesssim 0.1$, the influence of a magnetic field on the temperature relaxation can be ignored. For the intermediate magnetic field strength $0.1 < \Omega /\nu^T_{ei} < 1$ both magnetic field and nonlocal effects are important and their interplay defines the relaxation of the hot spot temperature.

\section{Conclusions}\label{sec8}

We have applied a systematic closure procedure \cite{bychenkov} to a magnetized plasma in the regime of moderate collisionality where $k_\parallel \lambda_{ei}<1$ and $k_\perp < Max \{ 1/\rho, 1/\lambda_{ei}\}$. We have derived nonlocal hydrodynamic equations which are valid for small perturbations about an equilibrium homogeneous state and for a plasma with high ionic charge. Our transport theory is fully equivalent to the linearized electron kinetic equation but possesses advantages of the reduced model which describes plasma in terms of just few moments of the distribution function. The closure procedure provides nonlocal relations between electron fluxes
and gradients in the plasma density and electron temperature, and an electrostatic field. These tensor transport relations account for all orientations of the magnetic field and the plasma gradients: longitudinal, oblique, and transverse. 

Competition between nonlocal and magnetic field effects is poorly understood and our theory provides a rigorous means of examining the relative importance of both factors. We have found that large magnetic fields contribute to the enhancement of electron conductivity in the weakly collisional regime. This increase can offset inhibition due to nonlocal effects. This is an important effect that is necessary to explain transport phenomena in laser produced plasmas \cite{nicolai}. The practical importance of our theory lies in providing benchmark results for comparisons with the phenomenological models such as Ref. \cite{nicolai} or future kinetic simulations (cf. Refs. \cite{brunner,brunner1}). 

We have described the application of nonlocal hydrodynamics to the MHD wave dispersion and damping calculations and to hot spot temperature relaxation. There are important enhancements to collisional damping of MHD modes in the nonlocal transport regime. The presence of a strong magnetic field is the main source of thermal transport inhibition in the relaxation of a hot spot. For the intermediate values, $0.1 < \Omega /\nu^T_{ei} < 1$, the hot spot displays the simultaneous effects of electron gyro motion and nonlocal transport. 

Our results provide a next step in the ongoing efforts to develop a plasma hydrodynamics model which is valid for the wide range of plasma conditions and includes nonlinear response to large amplitude perturbations. The application of linear theory to experimental data, such as \cite{montgomery,nicolai,ditmire,glenzer1}, and comparison with kinetic simulations and its generalization to include local dependence on plasma parameters, as in Ref. \cite{batishchev}, could lead to a nonlocal nonlinear transport theory in magnetized plasmas. Linear nonlocal hydrodynamics models provide a description of the plasma thermal response to electromagnetic radiation in stimulated Brillouin scattering, filamentation and in the filament resonant instability \cite{sbs-fil,sf,inst}. Current results will allow generalization of these processes to the case of magnetized plasmas.

\begin{acknowledgments}
A. V. B. and W. R. acknowledge generous support from the Alberta Ingenuity Fund. This work was partly supported by the Natural Sciences and Engineering Research Council of Canada and the Russian Foundation for Basic Research (grant N 03-02-16428).
\end{acknowledgments}

\appendix
\section{Nonlocal transport coefficients}
The components of the electrical conductivity tensor, {\boldmath
$\hat{\sigma}$}, and thermocurrent tensor, {\boldmath $\hat{\alpha}^j$} are presented in the following form:
\begin{eqnarray}
&& \sigma_{11}=\sigma_{\perp}^0-\sigma_{\perp}+\sigma_{\perp}^{E_1} \frac{k_{\parallel}^2}{k^2} \,, \quad \sigma_{12}=\sigma_{\wedge}^0-\sigma_{\wedge}-\sigma_{\perp}^{E_2} \frac{k_{\parallel}^2}{k^2} \,, \nonumber \\
 && \sigma_{21}=\sigma_{\wedge}^0-\sigma_{\wedge}+\sigma_{\wedge}^{E_1} \frac{k_{\parallel}^2}{k^2} \,,\quad \sigma_{22}=\sigma_{\perp}^0-\sigma_{\perp}+\sigma_{\wedge}^{E_2} \frac{k_{\parallel}^2}{k^2} \,, \nonumber \\
&& \alpha_{11}=\alpha_{\perp}^0-\alpha_{\perp}^q+\alpha_{\perp}^{E_1} \frac{k_{\parallel}^2}{k^2} \,, \quad \alpha_{12}=\alpha_{\wedge}^0-\alpha_{\wedge}^q-\alpha_{\perp}^{E_2} \frac{k_{\parallel}^2}{k^2} \,,  \\
 && \alpha_{21}=\alpha_{\wedge}^0-\alpha_{\wedge}^q+\alpha_{\wedge}^{E_1} \frac{k_{\parallel}^2}{k^2} \,,\quad \alpha_{22}=\alpha_{\perp}^0-\alpha_{\perp}^q+\alpha_{\wedge}^{E_2} \frac{k_{\parallel}^2}{k^2} \,, \nonumber \\
&& \alpha_{31}=\alpha_{\parallel}^0-\alpha_{\parallel}^q-\alpha_{\parallel}^{E_1}
\frac{k_{\parallel}^2}{k^2} \,,\quad \alpha_{32}=\alpha_{\parallel}^{E_2}
\frac{k_{\parallel}^2}{k^2}\,\,. \nonumber
\end{eqnarray}
The coefficients $\sigma_a^0$ and $\alpha_a^0$ are given by the transport theory for strong collisions (\ref{Braginskii}) and the introduced new transport coefficients, such as
\begin{eqnarray}
&& \sigma_a = \frac{e^2 n_e}{T_e k^2} \frac{D_{aT}^{NT}}{D_{NT}^{NT}}\,, \qquad \sigma_a^{E_1} =  \frac{e^2 n_e}{T_e k^2} \left[  J_a^{E_1} - \frac{D_{NT}^{E_1T}}{D_{NT}^{NT}}J_a^N - \frac{D_{TN}^{E_1N}}{D_{NT}^{NT}}J_a^T \right]\,,\nonumber \\ \label{tr_coef}
&& \alpha_a^j = \frac{e n_e}{T_e k^2} \frac{D_{aT}^{NT}-D_{aN}^{TN}}{D_{NT}^{NT}} \,, \quad \sigma_a^{E_2} =  \frac{e^2 n_e}{T_e k^2} \left[  J_a^{E_2} - \frac{D_{NT}^{E_2T}}{D_{NT}^{NT}}J_a^N-
\frac{D_{TN}^{E_2N}}{D_{NT}^{NT}}J_a^T \right]
\\&& \alpha_a^q = \frac{e n_e}{T_e k^2} \frac{D_{q_aT}^{NT}}{D_{NT}^{NT}}\,, \qquad \alpha_a^{E_1} =  \frac{e n_e}{T_e k^2} \left[  J_{q_a}^{E_1} - \frac{D_{NT}^{E_1T}}{D_{NT}^{NT}}J_{q_a}^N - \frac{D_{TN}^{E_1N}}{D_{NT}^{NT}}J_{q_a}^T \right]\,,\nonumber \\
&& \chi_a = \frac{n_e}{k^2} \frac{D_{{q_a}T}^{NT}-D_{q_aN}^{TN}}{D_{NT}^{NT}} \,,
\quad \alpha_a^{E_2} =  \frac{e n_e}{T_e k^2} \left[  J_{q_a}^{E_2} -
\frac{D_{NT}^{E_2T}}{D_{NT}^{NT}}J_{q_a}^N-
\frac{D_{TN}^{E_2N}}{D_{NT}^{NT}}J_{q_a}^T \right] \,,\nonumber
\end{eqnarray}
Expressions in (\ref{tr_coef}) can be calculated in terms of moments of the basic distribution functions satisfying Eq. (\ref{basic}).

\section{Dispersion relation}

This appendix presents the result of calculation of the dispersion relation for MHD waves. We write the nonlocal hydrodynamic equations (\ref{hydro1}) and the wave equation (\ref{wave}) in the matrix form. The vanishing of the following determinant leads to the dispersion relation.
\begin{eqnarray}\nonumber
 \left | \begin{array}{ccccccc}
{k}^2 \sigma & k_{\perp}\sigma_{\perp} & -k_{\perp}\sigma_{\wedge} & i {k}^2
\alpha^j &0&0&0 \\ {k}_{\perp} \sigma_{\perp} &
\sigma_{\perp}\!\!+\displaystyle{\frac{k_{\perp}^2}{k^2}} \sigma_{11} &
-\sigma_{\wedge}\!\!-\sigma_{12} & i {k}_{\perp}^2 \alpha_{\perp}^j & -1&0&0 \\
-{k}_{\perp}^2 \sigma_{\wedge} &
\sigma_{\wedge}\!\!+\displaystyle{\frac{k_{\perp}^2}{k^2}} \sigma_{21} &
\sigma_{\perp}\!\!+\sigma_{22} & -i {k}_{\perp}^2 \alpha_{\wedge}^j & 0&-1&0 \\
{k}^2 \alpha T_e & k_{\perp}\alpha_{\perp}^j T_e & - k_{\perp} \alpha^j_{\wedge}T_e
& i {k}^2 \chi \! + \!\! \displaystyle{\frac{5 n_e \omega}{2}} & 0&0&\omega n_e T_e
\\ 0&-\omega k_{\parallel}^2 c^2/4 \pi&0&0& \!\!\!\! i (\omega^2\!\!-k_{\parallel}^2
v_{A}^2)&0&0  \\0&0&-\displaystyle{\frac{k^2 c^2 \omega^2}{4 \pi}} &0&0& \!\!\!\! i (\omega^2\!\!-k^2 v_{A}^2) & k_{\perp} \Omega_i \displaystyle{\frac{k^2 c^2 T_e}{4
\pi e}} \\0&0&0&-\omega^2 /T_e &0&i k_{\perp} \displaystyle{\frac{\Omega_i}{e
n_e}}&k^2 c_s^2-\omega^2 \end{array}
 \right |
\end{eqnarray}
where $\{ \sigma, \alpha, \chi \} =k^{-2}\{k_{\parallel}^2
\sigma_{\parallel}+k_{\perp}^2 \sigma_{\perp}, \,k_{\parallel}^2
\alpha_{\parallel}+k_{\perp}^2 \alpha_{\perp}, \,k_{\parallel}^2\chi_{\parallel}
+k_{\perp}^2 \chi_{\perp} \}$, $c_s = \sqrt{Z T_e /m_i}$ is the ion-acoustic
velocity, and $v_A = B/\sqrt{4 \pi m_i n_i}$ is the Alfven velocity.

This dispersion relation could be rewritten in the form (\ref{deq}) where the
parameters $A$, $B$, $C$, $D$, and $E$, which are the functions of the nonlocal transport coefficients arise naturally. They are
\begin{eqnarray}\nonumber
A &=& n_e (k_A^2 \omega \sigma_3 + i \omega^2 k^2 \sigma_- )/2\,, \\ \nonumber B &=&
n_e \omega^2 ( k^2 \omega \sigma  - i k_A^2 \sigma_- )/2, \\ \nonumber C &=& k_A^2
k^2 (\chi \sigma_- -T_e \alpha_-) + i \omega^2 k^4 (\chi \sigma - T_e \alpha^2), \\
\nonumber D &=& \omega k^4 (\chi \sigma_+ -T_e \alpha_+) - i k_A^2 k^2 (\chi
\sigma_3 - T_e \alpha_4),\\ \nonumber E &=& \frac{T_e}{e} k_{\perp}^2 \omega (k^2
(\sigma \alpha_{\wedge}-\alpha \sigma_{\wedge}) + i k_A^2 \omega \alpha_3) \,,
\end{eqnarray}
where we introduced the following combinations of the transport coefficients:
\begin{eqnarray}\nonumber
\sigma_- &=& \sigma_{\parallel} \sigma_{\perp} + \sigma \sigma_{11}\,, \qquad \sigma_+ = \sigma (\sigma_{\perp}+\sigma_{22})+ \frac{k_{\perp}^2}{k^2} \sigma_{\wedge}^2, \\ \nonumber
\alpha_- &=&( \frac{k_{\parallel}^2}{k^2} \alpha_{\parallel}^2 \sigma_{\perp}+ \frac{k_{\perp}^2}{k^2} \alpha_{\perp}^2 \sigma_{\parallel}+ \alpha^2 \sigma_{11}), \\ \nonumber
\alpha_+ &=& \alpha^2 (\sigma_{\perp}+\sigma_{22})+ \frac{k_{\perp}^2}{k^2}(2 \alpha_{\perp} \alpha_{\wedge} \sigma_{\wedge} - \alpha_{\wedge}^2 \sigma), \\ \nonumber
\alpha_3 &=&(2 \sigma_{\parallel}(\sigma_{\wedge}\alpha_{\perp}-\sigma_{\perp} \alpha_{\wedge})+ 2 \sigma_{11} (\sigma_{\wedge}\alpha -\sigma \alpha_{\wedge})+ (\sigma_{12}+\frac{k_{\parallel}^2}{k^2}\sigma_{21})(\sigma_{\parallel}\alpha_{\perp}-\sigma_{\perp} \alpha_{\parallel})), \\ \nonumber
\alpha_4 &=&\frac{k_{\perp}^2}{k^2} \alpha_{\wedge}(als+\sigma_-)+(\sigma_{\perp}+\sigma_{22})\alpha_- +(\alpha^2 \sigma_{12} \sigma_{21} + \frac{k_{\parallel}^2}{k^2} \alpha_{\parallel}^2 \sigma_{\wedge}^2+ \alpha \alpha_{\parallel} \sigma_{\wedge} (\sigma_{12}+\frac{k_{\parallel}^2}{k^2}\sigma_{21})), \\ \nonumber
\sigma_3 &=& (\sigma_{\parallel}(\sigma_{\perp}^2 + \sigma_{\wedge}^2)+\sigma(\sigma_{11}(\sigma_{\perp}+\sigma_{22})+ \sigma_{12}\sigma_{21})+  \sigma_{\wedge}\sigma_{11} (\sigma_{12}+\frac{k_{\parallel}^2}{k^2}\sigma_{21}))+ \\
&& \sigma_{22}\sigma_{\parallel}\sigma_{\perp}+\frac{k_{\perp}^2}{k^2} \sigma_{11} \sigma_{\wedge}^2 \,. \nonumber
\end{eqnarray}
The derived dispersion relation (\ref{deq}) describes small amplitude MHD modes in the $k_\parallel \lambda_{ei}<1$--range of the wave numbers. This encompasses the range of wave numbers from the adiabatic perturbations for small $k$ to the isothermal perturbations for shorter wavelength.

\vspace*{35mm}\begin{center}{\large \bf Figure captions}\end{center}
\begin{figure}[!ht]
\caption{Longitudinal electrical conductivity normalized to $\sigma_{0}=32 e^2 n_e/3 \pi m_e \nu_{ei}^T$ for $k_\perp/k_\parallel=1000$ and $\Omega / \nu^T_{ei} = 0.01$ (1), $0.1$ (2), $0.3$ (3), $1.$ (4), $3$ (5), and $10$ (6).} \label{fig1}
\end{figure}

\begin{figure}[!ht]
\caption{Longitudinal temperature conductivity normalized to $\chi_{0}=(200/3 \pi) n_e v_{Te} \lambda_{ei}$ for $k_\perp/k_\parallel=1000$ and $\Omega / \nu_{ei} = 0.01 $(1), $0.1$ (2), $0.3$ (3), $1.$ (4), $3$ (5), and $10$ (6).} \label{fig2}
\end{figure}

\begin{figure}[!ht]
\caption{Longitudinal electrical conductivity normalized to $\sigma_{0}=32 e^2 n_e/3 \pi m_e \nu_{ei}^T$ for $k_\perp=k_\parallel$ and $\Omega / \nu^T_{ei} = 0.01 $(1),$ 0.1$ (2), $0.3$ (3), and $10$ (4).} \label{fig3}
\end{figure}

\begin{figure}[!ht]
\caption{Longitudinal temperature conductivity normalized to $\chi_{0}=(200/3 \pi) n_e v_{Te} \lambda_{ei}$ for $k_\perp=k_\parallel$ and $\Omega / \nu_{ei} = 0.01$ (1), $0.1$ (2), $0.3$ (3), and $10$ (4).}
 \label{fig4}
 \end{figure}

\begin{figure}[!ht]
\caption{Transversal electrical conductivity normalized to $\sigma_{0}=32 e^2 n_e/3 \pi m_e \nu_{ei}^T$ for $k_\perp=k_\parallel$ and $\Omega / \nu^T_{ei} = 0.01 $(1),$0.1$ (2), $0.3$ (3), and $1$ (4).}
 \label{fig5}
 \end{figure}

\begin{figure}[!ht]
\caption{Transversal temperature conductivity normalized to $\chi_{0}=(200/3 \pi) n_e v_{Te} \lambda_{ei}$ for $k_\perp=k_\parallel$ and $\Omega / \nu_{ei} = 0.01 $ (1), $0.1$ (2), $0.3$ (3), $1$ (4) and $3$ (5).}
 \label{fig6}
 \end{figure}

\begin{figure}[!ht]
\caption{ Oblique conductivity normalized to $e^2 n_e/m_e \nu_{ei}^T$ for
$k_\perp=k_\parallel$ and $\Omega / \nu^T_{ei} = 0.01 $(1), $0.1$ (2), $0.3$ (3), $1.$ (4), $3$ (5), and $10$ (6).} \label{fig7}
\end{figure}

\begin{figure}[!ht]
\caption{Oblique temperature conductivity normalized to $n_e v_{Te} \lambda_{ei}$ for $k_\perp=k_\parallel$ and $\Omega / \nu^T_{ei} = 0.01 $(1), $0.1$ (2), $0.3$ (3), $1.$ (4), $3$ (5), and $10$ (6). } \label{fig8}
\end{figure}

\begin{figure}[!ht]
\caption{The ratio of the transversal temperature conductivity to the longitudinal temperature conductivity for $k \lambda_{ei} = 0.1$ (large dots) and $0.3$ (small dots). Solid line corresponds to the local Braginskii's theory.}
 \label{fig9}
\end{figure}

\begin{figure}[!ht]
\caption{Solution (dots) to the dispersion equation for the MHD waves for real
($\omega$) and imaginary ($\gamma$) parts of the frequency (panels $a$ and $b$, correspondingly) in comparison with the result of conventional theory (solid lines) for a plasma with $p = 50$ and $h=10$.} \label{fig10}
\end{figure}

\begin{figure}[!ht]
\caption{Solution (dots) to the dispersion equation for the MHD waves for real
($\omega$) and imaginary ($\gamma$) parts of the frequency (panels $a$ and $b$, correspondingly) in comparison with the result of conventional theory (solid lines) for a plasma with $p = 4$ and $h=50$ } \label{fig11}
\end{figure}

\begin{figure}[!ht]
\caption{The temperature ($a$) and heat flux ($b$) profiles (dots) at the time $t=
2/ \nu^T_{ei}$ in comparison with the results of nonlocal theory without magnetic
field \protect{\cite{senecha}} (solid lines) and classical local approach
\protect{\cite{epperlein}} (dashed lines). The plasma parameters are: $\Omega / \nu^T_{ei} = 0.3$, $R/\lambda_{ei} =3$, and $Z=5$.} \label{fig12}
\end{figure}

\begin{figure}[!ht]
\caption{The temperature relaxation time as a function of the hot spot radius (dots) in comparison with the nonlocal theory without magnetic field
\protect{\cite{senecha}} (solid lines) and the classical local approach
\protect{\cite{epperlein}} (dashed lines) for different Hall parameters.}
\label{fig13}
\end{figure}

\newpage

\epsfig{figure=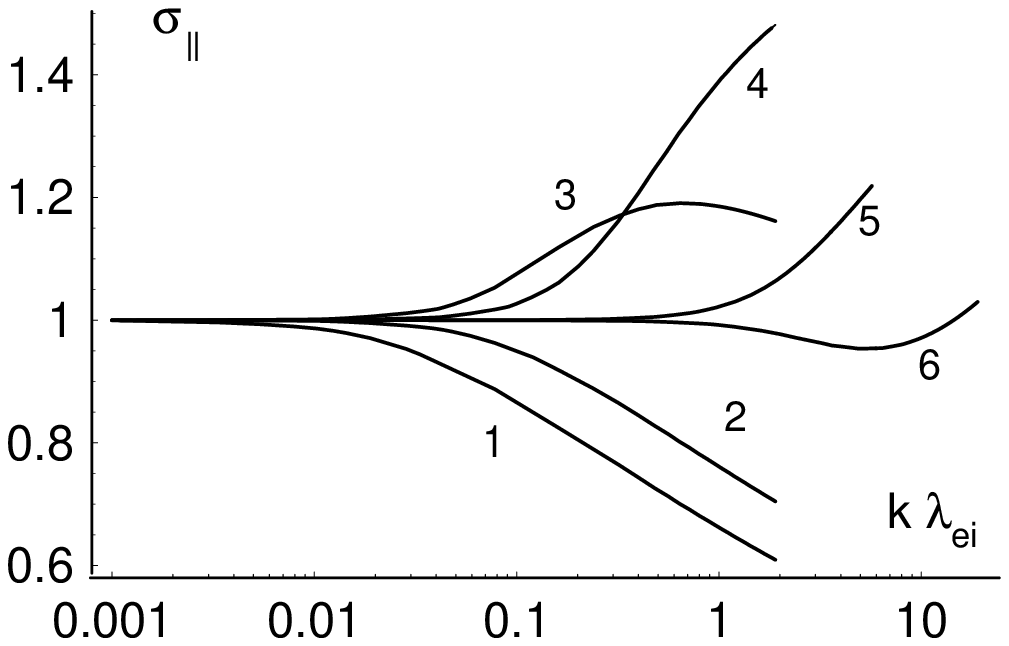, width=7.5cm}\\  \centerline{Figure \ref{fig1}}
\epsfig{figure=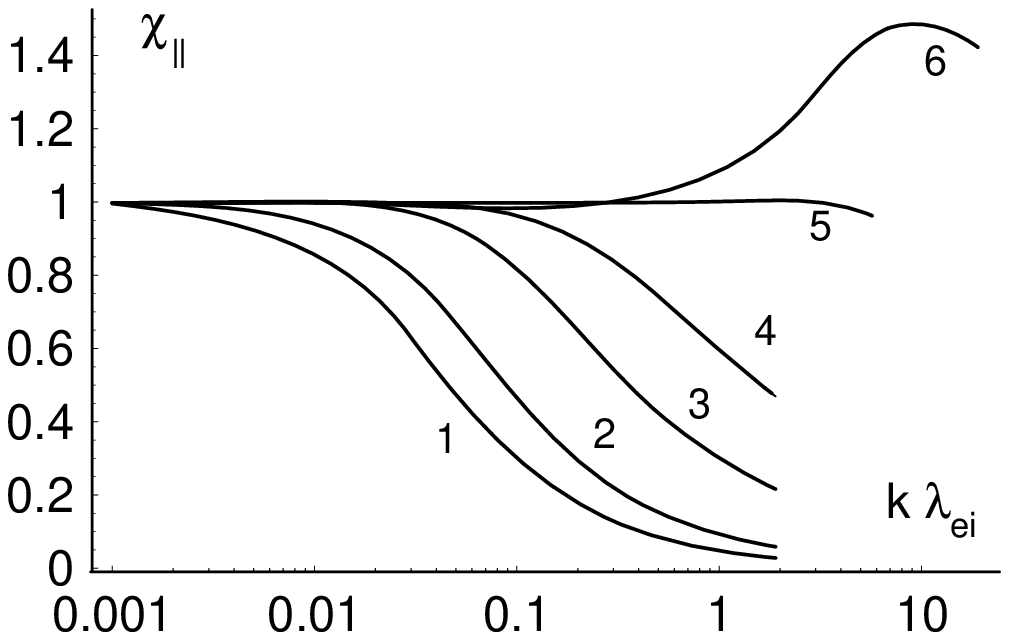, width=7.5cm}\\  \centerline{Figure \ref{fig2}}
\epsfig{figure=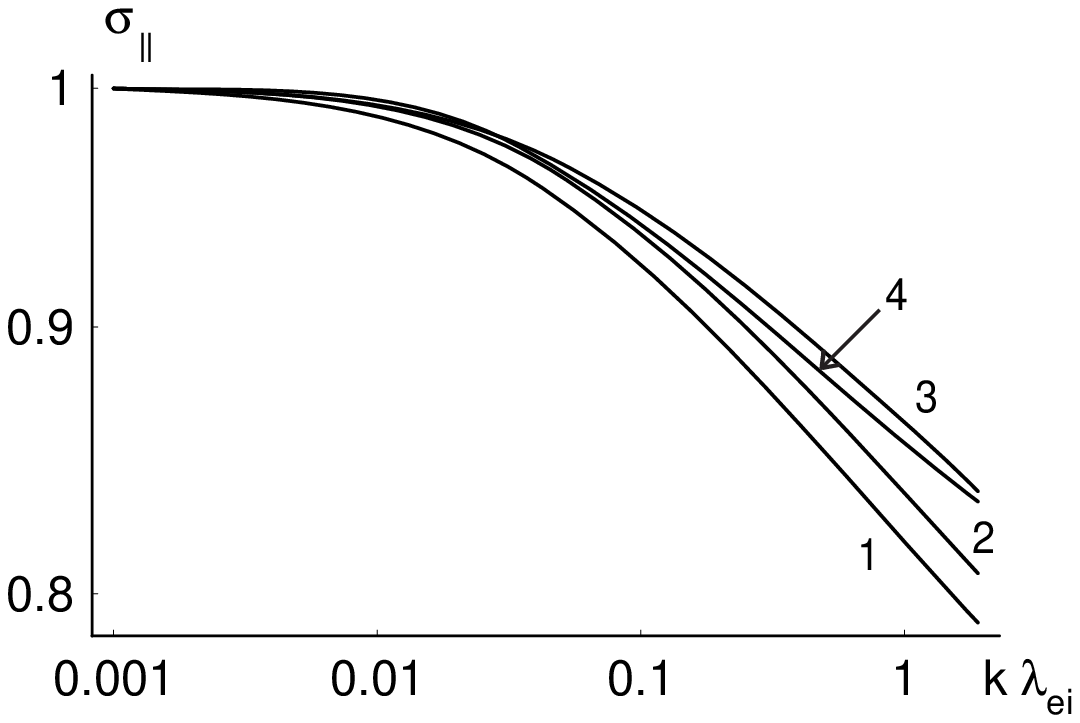, width=7.5cm}\\  \centerline{Figure \ref{fig3}}
\epsfig{figure=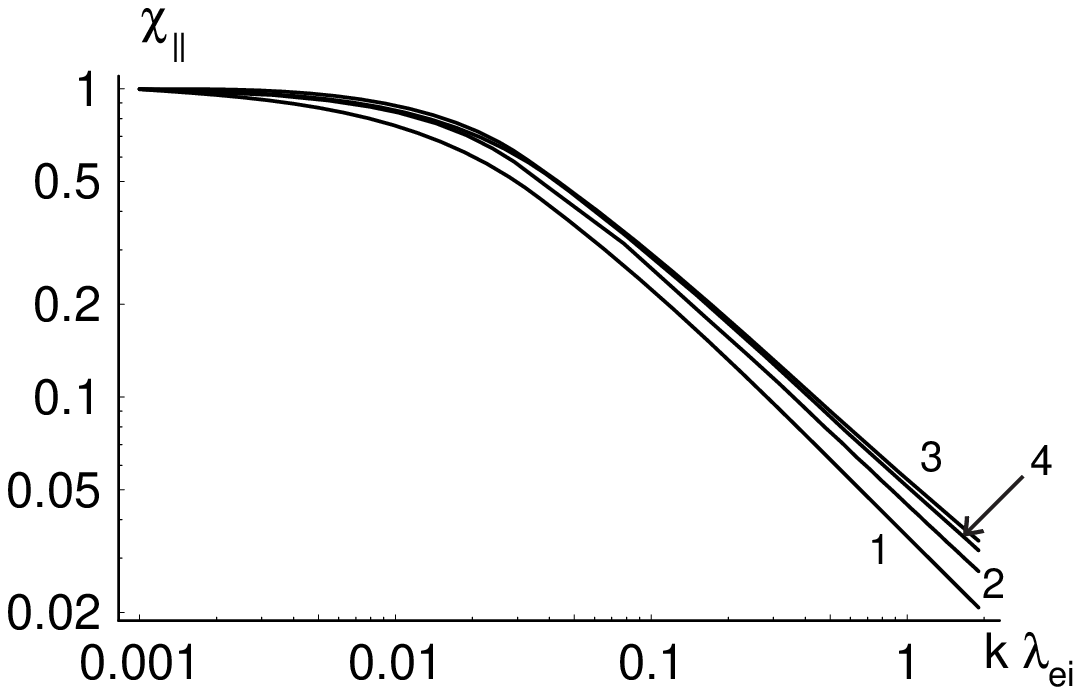, width=7.5cm}\\  \centerline{Figure \ref{fig4}}
\epsfig{figure=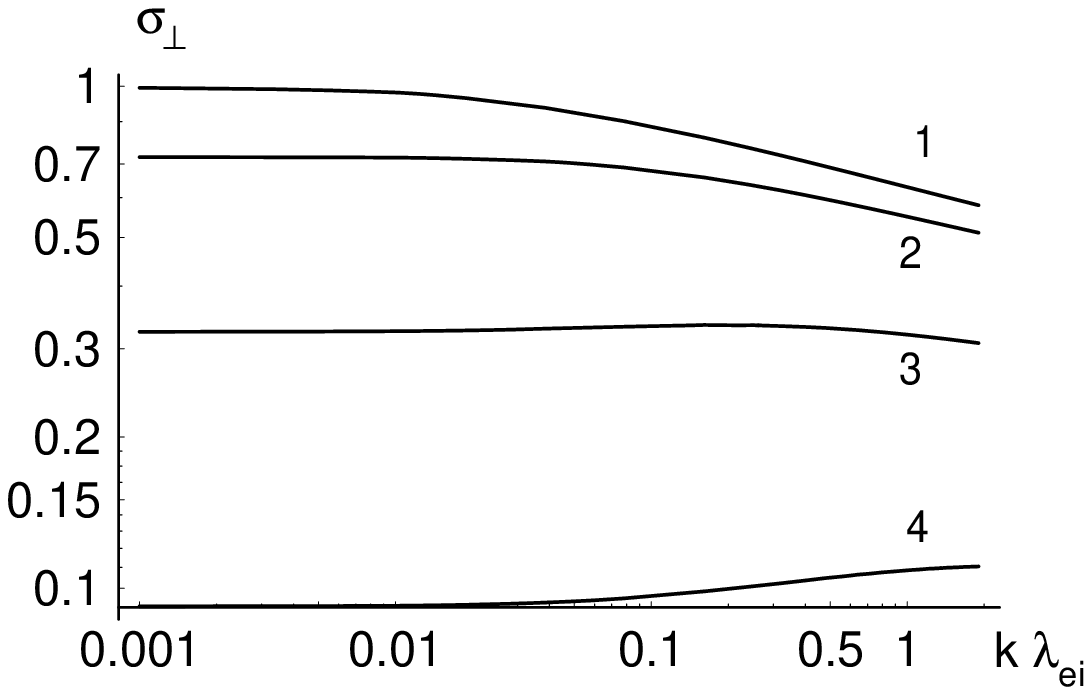, width=7.5cm}\\  \centerline{Figure \ref{fig5}}
\epsfig{figure=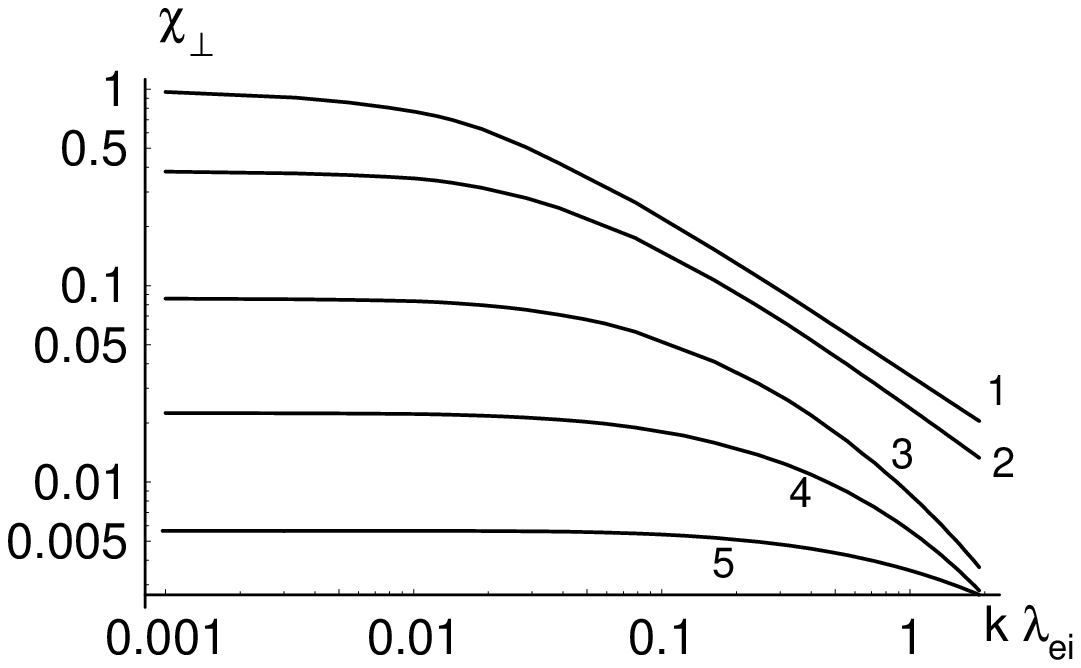, width=7.5cm}\\  \centerline{Figure \ref{fig6}}
\epsfig{figure=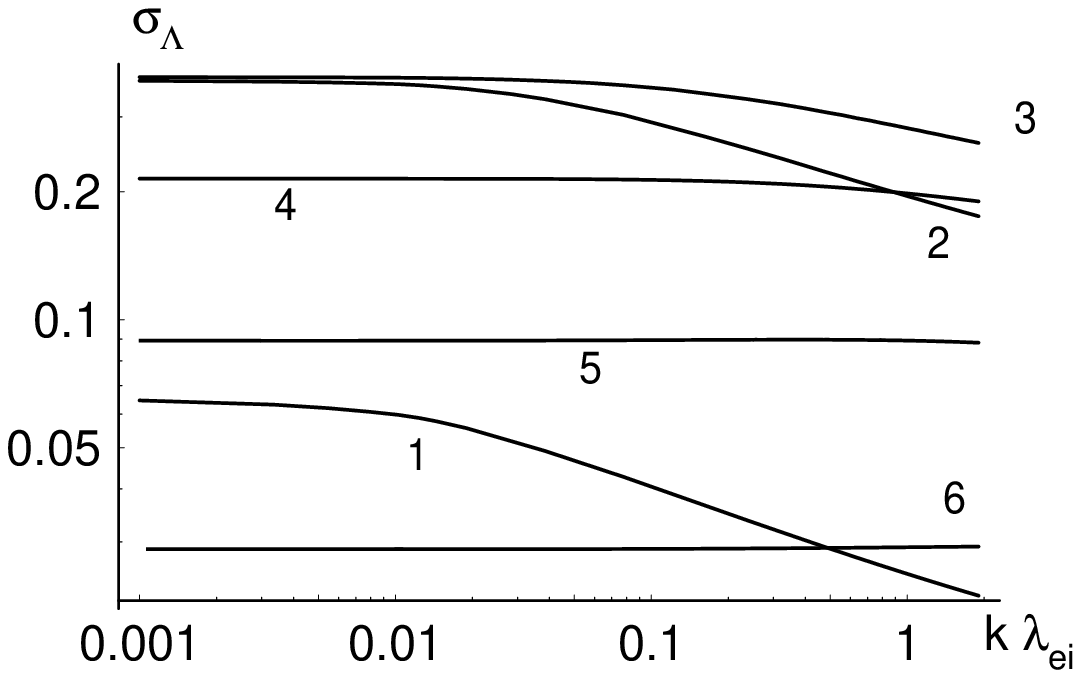, width=7.5cm}\\  \centerline{Figure \ref{fig7}}
\epsfig{figure=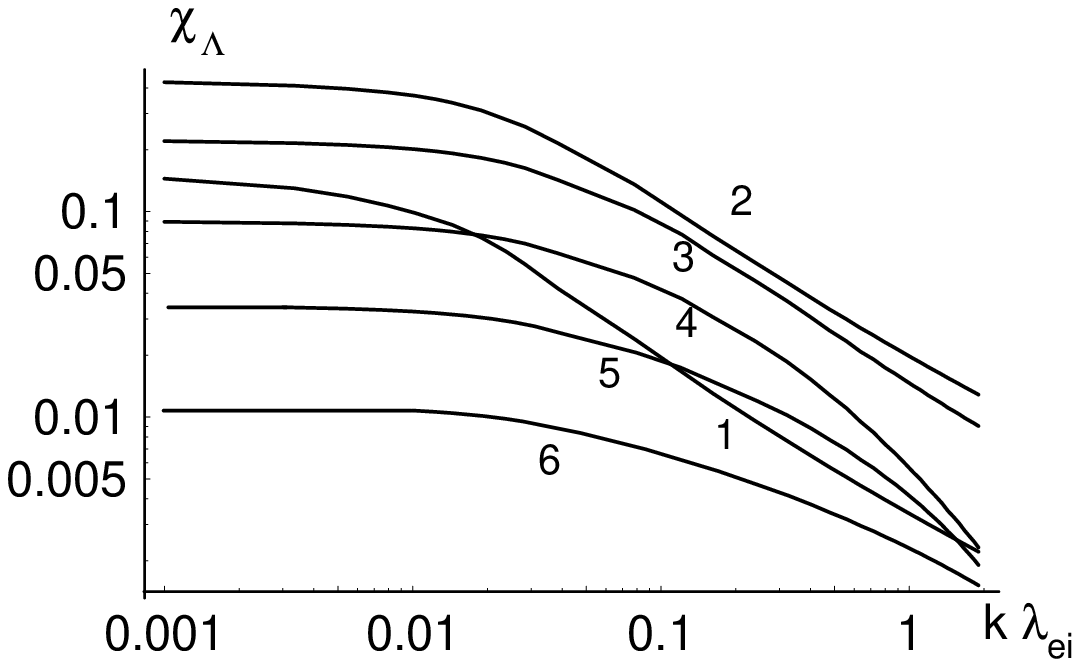, width=7.5cm}\\  \centerline{Figure \ref{fig8}}
\epsfig{figure=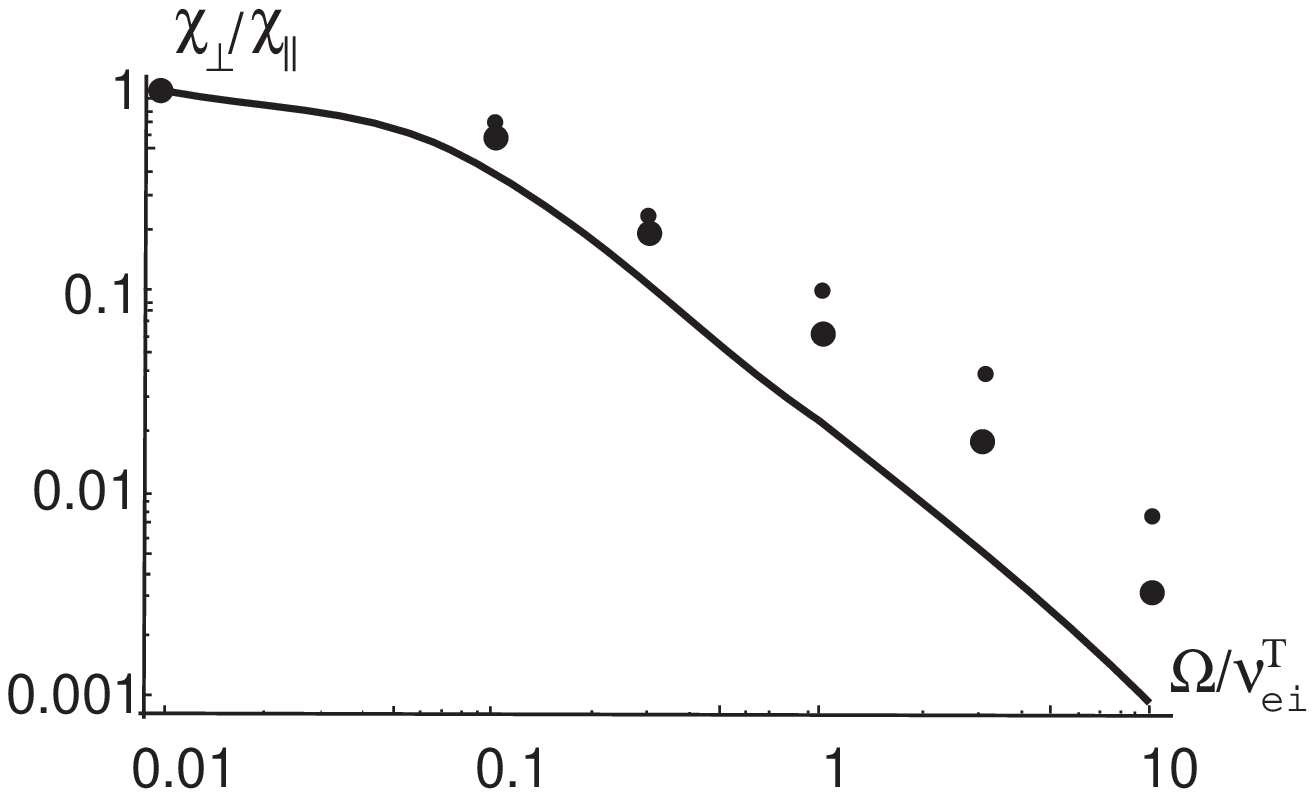, width=7.5cm}\\  \centerline{Figure \ref{fig9}}
\epsfig{figure=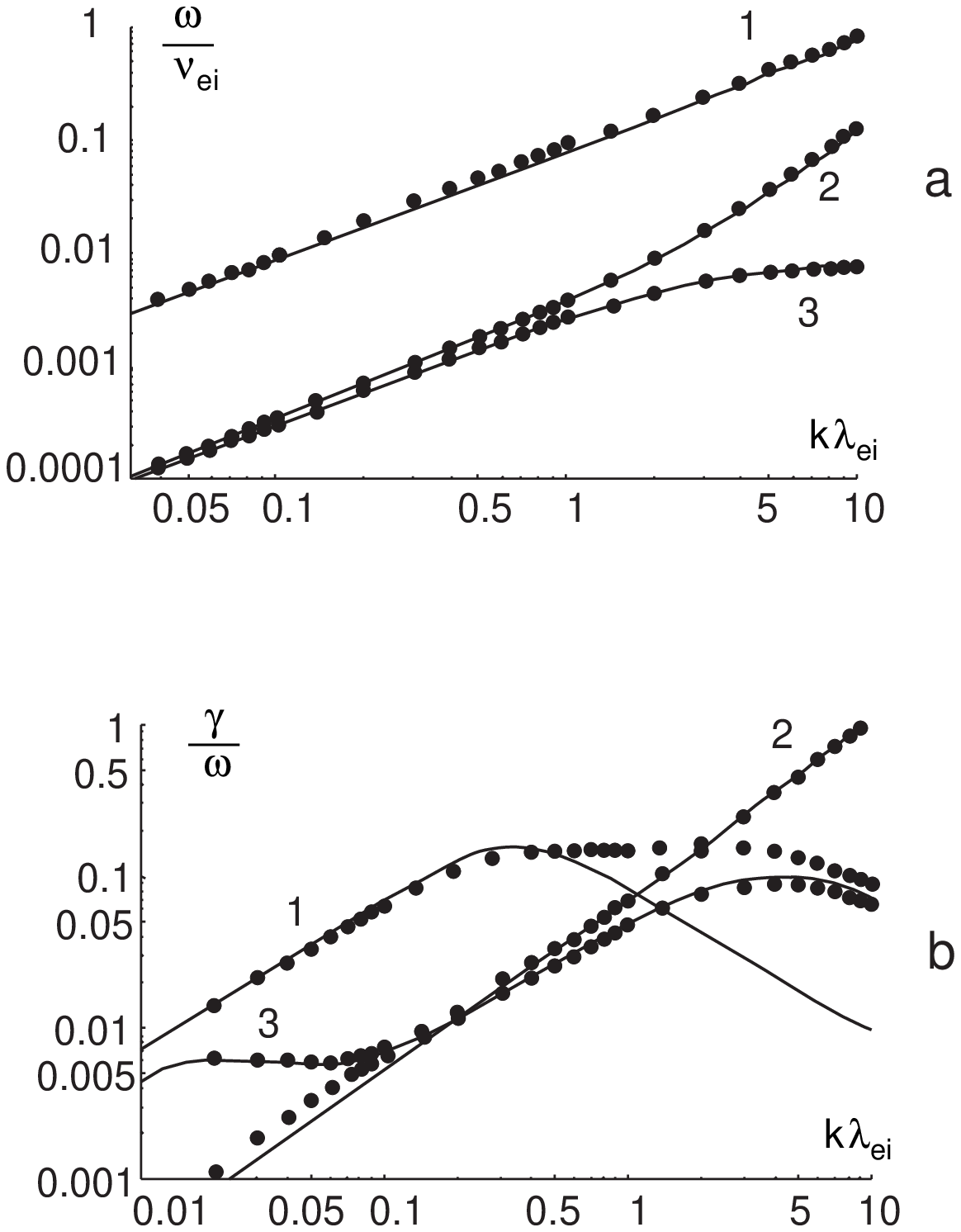, width=7.5cm}\\  \centerline{Figure \ref{fig10}}
\epsfig{figure=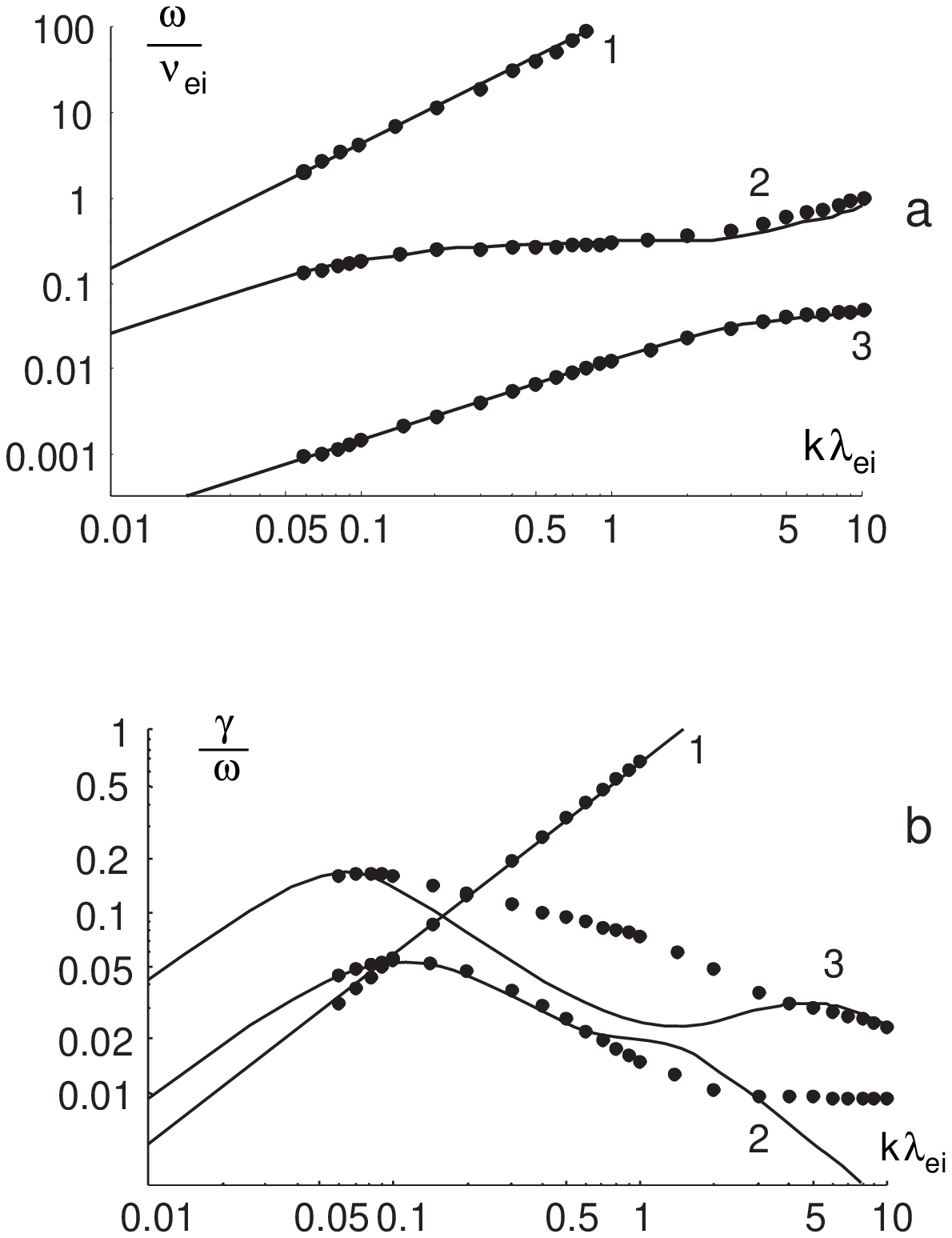, width=7.5cm}\\  \centerline{Figure \ref{fig11}}
\epsfig{figure=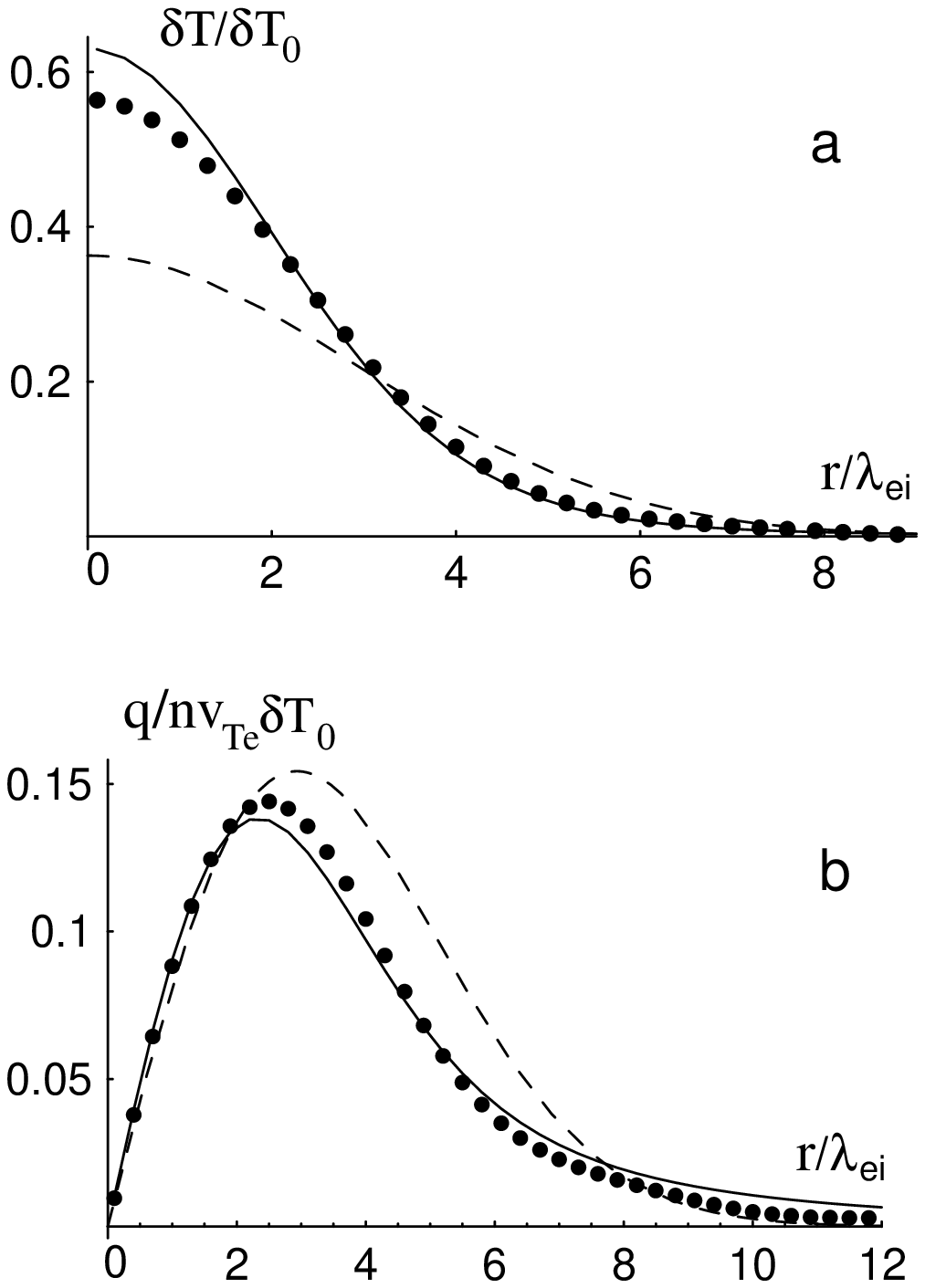, width=7.5cm}\\  \centerline{Figure \ref{fig12}}
\epsfig{figure=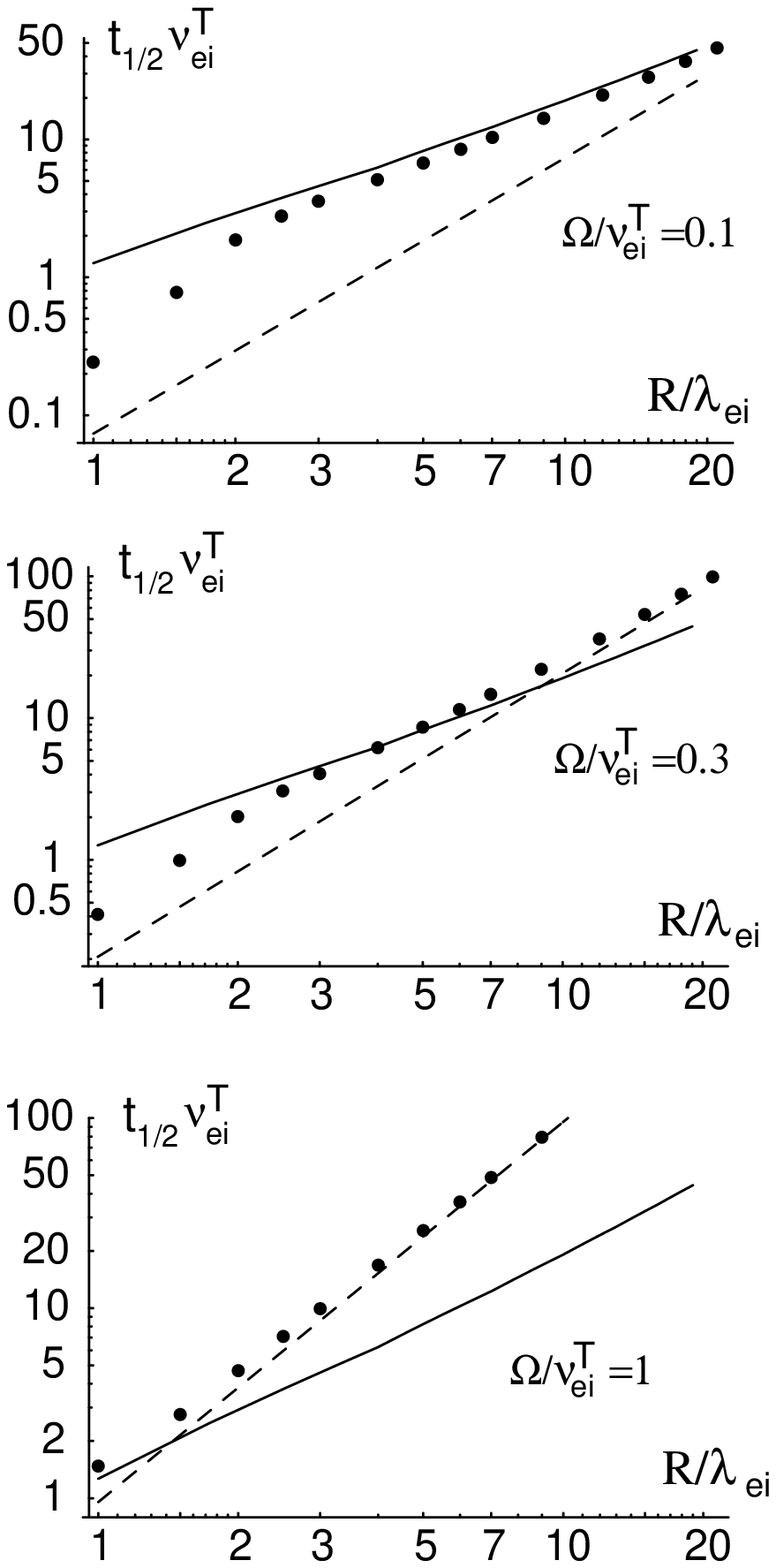, width=7.5cm}\\  \centerline{Figure \ref{fig13}}

\end{document}